\documentclass[aps,pre]{revtex4}    
\newcommand{\be}{\begin{equation}}
\newcommand{\ee}{\end{equation}} 
\newcommand{\ba}{\begin{eqnarray}}
\newcommand{\ea}{\end{eqnarray}}

\usepackage{epsfig}
\normalsize
\begin{document}
\title{Dynamics of Polydisperse Polymer Mixtures}
\author{I. Pagonabarraga$^{* \dag}$  and  M. E. Cates$^{*}$}
\affiliation{ $^{*}$Department of Physics and Astronomy, University of Edinburgh, JCMB Kings Buildings Mayfield Road, Edinburgh EH9 3JZ, Scotland
 $^{\dag}$ Departament de F\'{\i}sica Fonamental, Universitat de Barcelona \\
 Av. Diagonal 647, 08028-Barcelona, Spain}
\date{\today}
\begin{abstract}
We develop a general analysis of the diffusive dynamics of polydisperse polymers in the presence of chemical potential gradients, within the context of the tube model (with all species entangled). We obtain a set of coupled dynamical equations for the time evolution of the polymeric densities in a form proposed phenomenologically in recent work by N. Clarke, but with explicitly derived coefficients. For the case of chemical polydispersity (a set of chains that are identical except for having a continuous spectrum of enthalpic interaction strengths) the coupled equations can be fully solved in certain cases. For these we study the linearised mode spectrum following a quench through the spinodal, with and without a passive (polymeric) solvent. We also study the more conventional case of length polydisperse chains in a poor solvent. Here the mode structure is more complicated and exact analysis difficult, but enough progress can still be made to gain some qualitative insight. We briefly discuss the modifications required to allow for the presence of unentangled, low molecular weight species in the system. 
\end{abstract}
\maketitle
\section{Introduction}
There has been significant recent progress in the study of polydisperse
polymers under equilibrium conditions. This includes work on static phase 
coexistence (reviewed in ref 1) and on interfacial thermodynamics in 
situations of spatial inhomogeneity~\cite{el01}. Much of this work has 
been based on exploiting the fact that, within a standard Flory Huggins 
picture of the polymeric systems involved, the excess free energy depends 
only on a small number of moments of the distribution of polymer lengths or, 
if the chains are polydisperse not in length but in some other feature, 
moments of the distribution of that feature. An example of the latter is 
chemical polydispersity, where chains have, in effect, a continuous 
distribution of $\chi$ parameters~\cite{acp} but are otherwise identical; 
see also ref 3. (This can be realized experimentally by having a random 
copolymer in which the fraction of each type of monomer varies from chain 
to chain.)
This moment structure carries over into the nonlocal excess free energy 
for inhomogeneous states, at least in limiting cases~\cite{el01}; the 
moments become local moment density fields which depend nonlocally on 
the number densities of the individual polymeric species.  

Much less is understood about the dynamics of polydisperse systems. Consider, 
for example, the early stages of spinodal decomposition following a quench 
into the unstable regime. For length polydispersity this has been studied 
in recent papers by Clarke~\cite{clarke01} and by Warren~\cite{warren}, but 
even the qualitative picture remains far from clear. One complicating factor 
behind this is that, for the normal case of length polydispersity, the 
simplifications brought about by the moment structure in the excess free 
energy do not lead to 
any obvious simplification of the dynamical equations: the natural 
coordinates for dynamics are not the moment densities, but the species 
monomer densities.
(See ref 6 for a related discussion, in the context of colloids,
addressed there via a perturbation theory in the narrowness of the size
distribution.) Previous treatments of length polydispersity mapped the
polydisperse situation into that of a mixture with a finite set of
components by matching the lowest moments of the
distribution~\cite{olvera2}. For chemical polydispersity, however, there
is significantly more carry-over of the simplifications in the excess
free energy to dynamics. The basic reason for this is that such chains are all
dynamically degenerate (they have the same mobilities and structures)
even though they are subject to different thermodynamic forces. Hence
they can, in effect, be linearly recombined into moment densities, which
diagonalize the thermodynamics. This is not possible for length
polydispersity where the mobilities (and also the structure factors) of
the polymeric species vary from one chain to another; in general there
is then no practical way to simultaneously diagonalize the thermodynamic 
and the mobility-related factors in the diffusion matrix.  

	In this paper we develop a general formalism which includes, in principle, both chemical and length polydispersity (among others). In part this formalism resembles that of Clarke~\cite{clarke01} but it is more general and supplements his phenomenological equations with explicit calculation of the coefficients from a microscopic model based on the tube dynamics. The latter approach was developed previously in a similar context (but without polydispersity) by Brochard~\cite{Brochard}. These developments are presented in sections \ref{sec:basic} and \ref{sec:micro}. In section \ref{sect:funct} we present the free energy functional employed thereafter and in section \ref{sec:chempoly} we address in some detail the case of chemically polydisperse chains. These are considered in the melt and also in the presence of a passive solvent (one that does not have any direct enthalpic preference between the chemically polydisperse species). For the first case the dynamical equations can be solved in detail; in the 
second case, we find a relatively simple mode structure requiring a 
two-by-two diagonalization whose details involve the shape of the 
polydisperse distribution in the homogeneous (parent) phase. This simplifies 
further for the case of a symmetric parent.

These results are then contrasted with those in the literature for the case 
of length
polydispersity, which is studied further in section \ref{sect:length}. 
Although, for the reasons stated above, the same level of progress cannot 
be made in this case, the formal structure of the problem (which involves, 
for example, a coupling to negative moments of the length distribution) is 
rather interesting.
We are able to make some indicative progress by using a truncation scheme 
along the lines suggested by Warren~\cite{warren}, and indicate avenues for 
further analysis. Section \ref{sec:conclusions} contains our conclusions.

\section{System and basic dynamics}
\label{sec:basic}
We consider a polydisperse mixture of polymers. Each species is characterized 
by the chain number density $\rho_i({\bf r})$. (We use a discrete notation, 
but the continuum limit can be taken finally, as required.) The corresponding 
monomer density 
is written as $\phi_i({\bf r})=N_i \rho_i({\bf r})$, where $N_i$ is the 
polymerization index of species $i$; note that this  neglects terms 
 $O(\nabla^2)$ terms, see sec.\ref{sect:funct}. Depending on the kind of 
polydispersity, $N_i$ may be the same for different species. When dealing 
with polydisperse mixtures, 
the chain number density is the natural density to work with (as 
opposed to monomer density), because it characterizes the overall 
motion of the chain.

The continuity equation for polymer species $i$ is
\be
\frac{\partial \rho_i}{\partial t}=-\nabla\cdot\left(\rho_i {\bf v}_i \right)
\label{eq:cont}
\ee
where ${\bf v}_i$ is the mean velocity of species $i$, and $\rho_i{\bf v}_i = {\bf J}_i^t$ is the total flux of that species. We can rewrite 
this equation explicitly as a diffusion equation by introducing the barycentric 
velocity as a reference velocity. The barycentric velocity ${\bf v}^m$ is 
defined as 
\be
{\bf v}^m = \sum_i\phi_i{\bf v}_i
\ee
Since the polymer mixture is assumed to be incompressible, we have that at 
every point $\sum_i\phi_i=1$. (So far this does not exclude the possibility 
of a low molecular weight `solvent' among the species $i$.) This mass 
conservation property 
implies that ${\bf v}^m$ is solenoidal, i.e. $\nabla\cdot{\bf v}^m=0$. If 
we introduce the following time derivative
\be
\frac{d}{d t} \equiv\frac{\partial}{\partial t}+{\bf v}^m\cdot\nabla
\ee 
then the continuity equation, eq \ref{eq:cont}, can be rewritten as
\be
\frac{d\rho_i}{d t} =-\nabla\cdot{\bf J}_i
\label{eq:contphi}
\ee
which has the form of a diffusion equation, expressed in terms of the 
diffusive flux ${\bf J}_i$ of species $i$, which coincides with the total 
flux only when measured in the barycentric frame of reference: 
\be
{\bf J}_i\equiv {\bf J}^t_i - \rho_i {\bf v}^m=\rho_i({\bf v}_i-{\bf v}^m)
\ee

We could, of course,  have chosen any other reference velocity in place of
 ${\bf v}^m$, although in that case 
the continuity equation does not have the convenient form of 
eq \ref{eq:contphi}. An exception is the 
special case of mechanical equilibrium ($\nabla p = 0$); for this situation, 
the 
diffusive flux can be defined with respect to any reference velocity 
leading 
to the same expression for it. But in general, concentration gradients in 
our system will induce pressure gradients and thence fluid flow, which is 
handled by use of the barycentric frame. 

\subsection{Phenomenological model}

Following Clarke's phenomenological model~\cite{clarke01}, we  can start from 
the diffusive (Cahn-Hilliard) \\dynamics expressed in eq \ref{eq:contphi} 
and {\it assume} that the diffusive flux of species $i$ has the expression
\be
{\bf J}_i = \frac{M^0_i}{N_i}\nabla\mu_i+\rho_i{\bf v}^r
\label{eq:clarkeflux}
\ee
where $\mu_i$ is the chemical potential of species $i$, and $M_i^0$ is a
phenomenological mobility coefficient, and ${\bf v}^r$ is an unknown
reference velocity that is common to all polymer species. We can use the\\
incompressibility condition (local conservation of mass) to fix this
reference velocity because $\sum_iN_i{\bf J}_i=0$. The diffusive flux then 
simplifies to
\be
{\bf J}_i = \sum_kM_{ik}
\nabla\mu_k
\label{eq:fluxpheno}
\ee
 where we identify the mobility matrix 
\be
M_{ik}=\frac{M^0_i}{N_i}\delta_{ik}-\rho_iM^0_k
\label{eq:genmob}
\ee

This model does not provide explicit expressions for the mobilities $M_i^0$. It seems reasonable to assume that they are proportional to the density, suggesting the more general form
\be
M_{ik}=A_i\rho_i\delta_{ik}-\rho_i\rho_k B_{ik}
\label{eq:pheno}
\ee
where for length polydispersity $A_i$ and $B_{ik}$ will depend on $N_i$. In general they might also depend on any other polydisperse feature such as the fraction of monomers of given chemical type on chain of species $i$.
In the next section we develop a particular microscopic model in which the only dependences enter through the $N_i$ (so that the case of chemical polydispersity becomes especially simple). We will show that we can recover the generic form  for the diffusive fluxes, eq \ref{eq:fluxpheno}, although now with explicit expressions for the mobility coefficients. 

\section{Microscopic model (following Brochard)}
\label{sec:micro}

For a dense mixture of long chains, we will use the tube model to describe polymer dynamics, as applied to interdiffusion by Brochard~\cite{Brochard}. 
We assume that all the chains are dynamically similar in all respects apart from their lengths: that is, we ignore the effect of chemical species on entanglement length, mobility {\em etc.}. According to classical reptation theory (which assumes that each polymer moves through a tube  made of the network)~\cite{DoiEdwards}, the center of mass velocity ${\bf v}_i$ of polymer species $i$, is related to its curvilinear velocity $w_i$ (which determines its motion through the tube), as follows:
\be
{\bf v}_i = w_i \frac{{\bf h}_i}{L_i}
\label{eq:vcm}
\ee
where $L_i$ is the contour length of the tube, and ${\bf h}_i$ is the end-to-end polymer vector. Under \\near-equilibrium conditions, these are related by 
\be
L_i^2 = \frac{N_i}{N_e}{\langle h_i^2\rangle}
\label{eq:length}
\ee
where the entanglement parameter $N_e$ is related to the tube diameter~\cite{DoiEdwards}. Note that $N_e$ should be smaller than $N_i$ for all species present; for length polydispersity, this restricts the applicability of the present model to the case where all species, including any `solvent', are fully entangled.

In using eq \ref{eq:length} we assume that the polymer conformation is not significantly distorted from its equilibrium counterpart. Hence, the dynamics we will describe below will hold only for chemical potential gradients weak enough that one does not enter the non-Newtonian flow regime, where the flow response to pressure gradients becomes nonlinear. Using the previous expression, the curvilinear and center-of-mass velocities {\em measured relative to a static network} are then related through
\be
|{\bf v}_i|^2 = \frac{N_e}{N_i}w_i^2
\label{eq:vcmeq}
\ee
In practice, the network of entanglements is common to all the polymers, but need not 
be static. If the system moves under the action of an external field, this network will 
be characterized by a common velocity, ${\bf v}_t$, or `tube velocity'. 
Equation \ref{eq:vcmeq} should then be generalized to
\be
|{\bf v}_i-{\bf v}_t|^2 = \frac{N_e}{N_i}w_i^2
\label{eq:vcmeq2}
\ee
Following Doi and Onuki~\cite{doi92}, we may now deduce the (linearized) 
equations of motion by minimizing the energy dissipation rate of the system. 
This has two contributions: one related to the free energy variation, the 
second to the energy dissipation due to the friction between the polymers and 
the tube. We analyze the two terms in turn.

\subsection{Free energy rate contribution}
\label{sec:freeen}
Generically, the free energy of a polydisperse polymer mixture can be 
expressed as the sum of an ideal and an excess contribution:
\be
F=\int d{\bf r} f[\{\rho_i({\bf r})\}]=\int d{\bf r}\left\{ k_BT\sum_i \rho_i({\bf r}) \log\rho_i({\bf r}) + f^{ex}[{\rho_i(\bf r})]\right\}
\label{eq:free}
\ee
 where $f^{ex}$ is in general a non-local function of the chain-number density.
The free energy rate (which we define as the contribution to the dissipation 
of the system due to work done against thermodynamic gradients) can then be 
written as
\be
\frac{\partial F}{\partial t} = \int \frac{\partial f({\bf r})}{\partial t}\, d{\bf r} = -\int \mu_i({\bf r}) \nabla\cdot\left({\bf v}_i\rho_i\right)
\, d{\bf r}\label{eq:freerate}
\ee
where the chemical potential is defined in general as $\mu_i({\bf r})=\delta F/\delta \rho_i({\bf r})$ 
 and where use has been made of mass conservation, eq \ref{eq:cont}. This result is general, although in 
section \ref{sect:funct} we will discuss specific examples for the free energy
of polymer mixtures, expressed in terms of moments of the chain number 
densities.

\subsection{Tube dissipation rate}

The tube dissipation rate arises from the friction of the polymer against the  network during its motion. Hence the relevant polymer velocity is the 
curvilinear one, $w_i$. The dissipation is then assumed to be proportional 
to its square, and to the number of monomers of each polymer species. If we 
denote by $\xi_{0i}$ the microscopic friction constant of species $i$, we 
can write the tube dissipation rate as
\be
W=\int d{\bf r} \sum_i \frac{1}{2}\phi_i \xi_{0i} w_i^2=\int d{\bf r} \sum_i \frac{1}{2} \xi_i |{\bf v}_i-{\bf v}_t|^2
\label{eq:diss}
\ee
where we have used the equilibrium relation between the curvilinear and the 
center of mass velocity for polymer species $i$, eq \ref{eq:vcmeq2}, and 
introduced the polymer friction coefficient, 
\be
\xi_i=\xi_{0}\frac{N_i}{N_e}\phi_i = \xi_{0}\frac{N_i^2}{N_e}\rho_i
\label{eq:friction}
\ee
Since $\xi_{0i}$ is a microscopic friction coefficient, related to the interaction between a given monomer and the network, we assume it is the same for all the species,  i.e. $\xi_{0i}=\xi_0$. While true for pure length polydispersity, when chemical polydispersity is present this will not be strictly correct. (The resulting errors should, however, be quantitative not qualitative in nature, especially, for example, when the chemical differences involve only deuteration.)

Brochard's picture has allowed us to obtain explicit expression for the friction coefficients, derived from a {\em microscopic} model. In particular, it predicts a 
quadratic dependence of the friction coefficient on the polymerization index $N_i$ of species $i$.  Within our assumption of constant $\xi_{0i}$, it is only for length polydispersity that the mobility contributions will depend on polydispersity. For other situations such as chemical polydispersity, its influence on the diffusion coefficients will enter only through the thermodynamics of the system.

Following Brochard~\cite{Brochard} the tube velocity is now determined by requiring that the friction force acting on the network should balance. Such a force can be obtained from the dissipation rate $W$ taking its derivative with respect to the tube velocity. Requiring $\delta W/\delta {\bf v}_t=0$ determines the tube velocity
\be
{\bf v}_t = \frac{\sum_i\xi_i {\bf v}_i}{\sum_i\xi_i}
\label{eq:tubevel}
\ee
Note that, except for the important case of length polydispersity, the tube velocity always coincides with the barycentric velocity in our model. This coincidence is very useful to analyse the diffusion, because under these circumstances $\rho_i ({\bf v}_i -{\bf v}_t)$ is already the diffusive flux of species $i$ (and we obtain below an explicit expression for ${\bf v}_i -{\bf v}_t$). In this respect, length polydispersity plays a distinctive role from the dynamical point of view. 
\subsection{Total dissipation rate}

The total dissipation rate (essentially the entropy production rate) for our polymer 
mixture is now written as
\be
R=\int d{\bf r} \left\{ \sum_i \left[\frac{1}{2} \xi_i |{\bf v}_i-{\bf v}_t|^2-\mu_i \nabla\cdot\left(\rho_i {\bf v}_i\right)\right]-p\nabla\cdot\sum_i\phi_i{\bf v}_i\right\}
\ee 
where the third term is introduced to enforce incompressibility (cf. eq \ref{eq:contphi}). 
Here $p$, the associated Lagrange multiplier, corresponds to the pressure.

The polymer velocities are derived by minimizing $R$. Imposing $\delta R/\delta {\bf v}_i=0$ we 
get
\be
\xi_i ({\bf v}_i-{\bf v}_t)+\rho_i\nabla \mu_i+\phi_i\nabla p=0
\ee
Summing over all components (and imposing incompressibility, 
i.e. $\sum_i\phi_i({\bf r})=1$) we recover 
\be
\nabla p = -\sum_i\rho_i\nabla\mu_i
\ee
which is simply the Gibbs-Duhem equation (this shows that the Lagrange multiplier $p$ 
corresponds indeed to the thermodynamic pressure). Using this explicit expression, the 
velocity of polymer $i$ has the form
\be
{\bf v}_i = {\bf v}_t - \frac{\rho_i}{\xi_i} \nabla\mu_i+ \frac{\phi_i}{\xi_i}\sum_k \rho_k\nabla\mu_k
\ee
From this we can write the diffusive flux of species $i$ as
\be
{\bf J}_i =\rho_i\left({\bf v}_t-{\bf v}^m - \frac{\rho_i}{\xi_i} \nabla\mu_i+ \frac{\phi_i}{\xi_i}\sum_k \rho_k\nabla\mu_k\right) 
\label{eq:flux}
\ee
We can again use the incompressibility condition to obtain an explicit expression 
for ${\bf v}_t-{\bf v}^m$; imposing $\sum N_i {\bf J}_i = 0$  leads in this case to

\be
{\bf v}_t-{\bf v}^m =\sum_i\left\{\frac{\rho_i\phi_i}{\xi_i}-\rho_i\sum_k\frac{\phi_k^2}{\xi_k}\right\}\nabla\mu_i
\label{eq:diffvels1}
\ee

which shows that these two velocities are not independent of each other, and that their 
difference is proportional to the chemical potential gradients, but not simply to the pressure 
gradient (hence this difference is not caused by a departure from mechanical equilibrium).

As we have already mentioned, in the absence of length polydispersity, ${\bf v}_t - {\bf v}^m$ 
is zero within our model, because we assume $\xi_i\propto\phi_iN_i$. In the remaining case of 
length polydispersity, the difference can be expressed as
\be
{\bf v}_t-{\bf v}^m =\frac{1}{\alpha}\sum_i\left\{\frac{1}{N_i}-\rho\right\}\rho_i\nabla\mu_i
\label{eq:diffvels2}
\ee
where $\rho=\sum_i\rho_i$ is the overall local chain density, and $\alpha\equiv\xi_0/N_e$. 

\subsection{Result for the mobility matrix}
Substituting eq \ref{eq:diffvels1} into the expression for the diffusive flux, 
eq \ref{eq:flux}, we get
\be
{\bf J}_i = \sum_k\left\{-\frac{\rho_i}{\xi_i}\delta_{ik}+\rho_k\left(\frac{\phi_i}{\xi_i}
+\frac{\phi_k}{\xi_k}\right)-\rho_k\sum_j\frac{\phi_j^2}{\xi_j}\right\}\rho_i\nabla\mu_k
\label{eq:mik}
\ee
 This diffusive flux is proportional to the chemical potential gradients. It has the form 
of  eq \ref{eq:fluxpheno}, with the generalized mobility matrix, eq \ref{eq:pheno}.  
But in this case we have obtained an explicit expression for the mobility coefficients. As 
had been hypothesized, the mobility is proportional to the chain number density. However, 
for this model we have derived the dependence on the polymer length (which enters through 
the friction coefficients and which turns out to be nontrivial). We can rewrite 
eq \ref{eq:mik} in the same form as eq \ref{eq:pheno},
with the following choice of the mobility matrix:
\be
M_{ik}=-\frac{\rho_i\rho_i}{\xi_i}\delta_{ik}+\rho_i\rho_k\left(\frac{\phi_i}{\xi_i}+\frac{\phi_k}{\xi_k}\right)-\rho_i\rho_k\sum_j\frac{\phi_j^2}{\xi_j}
\ee
This shows that the phenomenological model recovers the correct functional dependence for the diffusive flux.  

Comparing with the expression introduced in eq \ref{eq:pheno}, our analysis based on the physical hypotheses of Brochard's picture~\cite{Brochard}, yields
\be
A=-\frac{\rho_i}{\xi_i}\;\;\;\;,\;\;\;B=-\frac{\phi_i}{\xi_i}+\frac{\phi_k}{\xi_k}+\sum_j\frac{\phi_j^2}{\xi_j}
\label{eq:AB}\ee
In the subsequent sections we analyze in detail several applications of this result, for different forms of the polydisperse polymer free energy.

As mentioned previously, our expression for the diffusive flux,
eq \ref{eq:mik}, has the same general form as that obtained by Clarke
~\cite{clarke01} for the case of length polydispersity. The main
conceptual difference is that, while Clarke expresses diffusive fluxes
as gradients of the polymer chemical potential ( relative to the solvent
chemical potential)  plus  a global convective contribution with respect to an unspecified common velocity, in our case we define diffusive fluxes with respect to the barycentric velocity. (We will come back to this comparison in section \ref{sect:length}.)
So far, in fact, we have not introduced any explicit solvent, although the model as presented above does permit one to enter it as one of the species $i$, so long as this `solvent' is itself made of chains that are long enough to be entangled. We return to the issue of solvent dynamics in the following sections.

\section{Free energy functional}
\label{sect:funct}

We now discuss the common structure of the free energy functional for all the systems we address below. 
For definiteness we consider polymer free energies that are derived from a specific nonlocal free energy functional proposed in ref 2. It has the form (setting $k_BT=1$)
\be
{\cal F} = \int d{\bf r} \left\{\sum_k\rho_k({\bf r})(\log(\rho_k({\bf r}))-1)+F^{ex}[\{m_n({\bf r})\}]\right\}
\label{eq:nonlocal}
\ee
where the excess free energy depends on a finite set of moments. These moments are locally varying quantities that depend nonlocally on the associated chain number densities through 
\be
m_n({\bf r})=\sum_k\omega_{nk}\int N_k \tilde{w}_{k}({\bf r}-{\bf r}')\rho_k({\bf r}') \,d{\bf r}
\label{eq:omegamoments}
\ee
where $\tilde{w}_{k}({\bf r}-{\bf r}')$ is related  to the structure factor of a polymer chain of species $k$ in a specific way~\cite{footgD}. The quantity $\omega_{nk}$ defines the $n$th moment in terms of the polydisperse variable; in all cases of interest here it is a simple power (e.g. $\omega_{nk} = N_k^n$ in the case of length polydispersity). 

The use of this functional, which contains the information on the internal 
 structure of the chains, will allow us to address issues such as the dependence of the initial unstable modes on the wave vector, in the case of a system undergoing spinodal decomposition. For small wavevectors the nonlocal kernel $\tilde\omega_k$ appropriate to a Gaussian chain is well approximated by the following gradient expansion~\cite{el01}
\be  
\tilde{w}_{k}({\bf r}-{\bf r}')=(1+\frac{N_k}{24}\nabla^2)
\delta({\bf r}-{\bf r}')\equiv W_{0k}\delta({\bf r}-{\bf r}')
\label{eq:gradientex}
\ee 
where the additional subscript $0$ denotes the small wavevector limit. The operator $W_{0k}$ will be used in subsequent sections. Note that the chosen normalisation, $\int \tilde{w}_k ({\bf r}) \, d{\bf r} =1$, means that factors $N_k$ appears beside it in eq \ref{eq:omegamoments}. Thus, in the case of length polydispersity, the zeroth nonlocal moment $m_0({\bf r})$ (which corresponds to choosing $\omega_{0k} = N_k^0 = 1$) in fact describes the local monomer concentration. For chemical polydispersity it is again the local monomer concentration, with no discrimination between monomers of different chemical species. (In the absence of a solvent species $m_0({\bf r})$ is unity everywhere, because of incompressibility.)

Note that although we use a specific form, eq \ref{eq:nonlocal}, for the
nonlocal contributions to the polymer free energy, it is chosen so as to
reduce to a standard Flory Huggins theory, in the limit of zero
wavenumber. This requirement fixes the form of the term
$F^{ex}[\{m_n({\bf r})\}]$ in eq \ref{eq:nonlocal}. Moreover, once the
gradient expansion eq \ref{eq:gradientex} is adopted, the nonlocality
we choose has a wider justification than the underlying model defined by
eq \ref{eq:omegamoments}. In particular, the same form for the
gradients could be obtained by the random phase
approximation~\cite{el01}.
  A non-local free energy functional will in principle require, for consistency, a non-local expression for the mobility matrix.  Such effects
have been considered previously in the literature \cite{footn}. However,
for simplicity, we will make in this paper the 
assumption that the mobility coefficients can be regarded as local, as is normal practice\cite{olvera,clarke01,doi92}.

\section{Chemical polydispersity}
\label{sec:chempoly}

Chemical polydispersity describes a set of polymers all with the same length, but a continuously varying 
interaction parameter. The normal example is a melt of random copolymers of equal length made of two 
monomers, e.g. deuterated and hydrogenated, that differ slightly in their enthalpic interaction. Within 
the model outlined above, we assume that this is the {\em only} difference. Each chain is then described 
by a polydisperse quantity $f_i$ obeying $-1 \le f \le 1$ which controls the proportion of the different 
monomers present along it; they have no correlation other than this, and within a mean field approximation 
the presence of variable $f$ is equivalent to a spectrum of $\chi$ parameters affecting different 
chains~\cite{footweak}. The moment densities $m_n({\bf r})$ are then defined by the choice
 $\omega_{nk} = f_k^n$; and within Flory Huggins theory, the excess free energy is simply
 $F^{ex} = -\chi m_1^2$~\cite{acp}. Here $\chi$ is the Flory parameter controlling the 
interaction between the two different monomers.

Next we treat this case of the chemically polydisperse melt. However, we also allow for a solvent to be present. Technically, this is already included in the preceding formalism so long as the solvent is considered as `just another species' within our incompressible mixture of many polymers. Indeed there would be nothing to prevent us developing the general formalism to address
this case without singling the solvent out for any special treatment at all.
But the fact that its length and chemistry will usually be different from all the remaining polymers implies that the solvent will have a distinct dynamical behavior, which it is convenient to handle separately. Moreover, because of \\ incompressibility, both the thermodynamics and the dynamics of the solvent are subservient to the remaining species and this will allow us to focus on the dynamics of the moments $m_n({\bf r})$ of the chemically  polydisperse chains alone. These dynamics will, of course, be different according to whether a solvent is present or not. 

Note that this treatment of the solvent does not release us from the assumptions of the tube model made in our earlier derivation of the diffusive fluxes. In particular, this means that the solvent itself must consist of chains long enough to be fully entangled (though, in the cases addressed below, much shorter than the remaining, chemically polydisperse, chains of primary interest). However, as we will 
discuss in section \ref{sect:length}, we can later on use the general form of the equations for the diffusive fluxes to guess how these may behave for an unentangled solution. 

As discussed in section \ref{sec:micro} above, a key simplifying feature for chemical polydispersity is that (within the approximations we have taken) the structure factors and mobilities of the chains are independent of the polydisperse feature $f_i$. In fact, many of the results described below for this case will also hold for any other kind of polydispersity with the same attribute, even if the structure of the excess free energy is not that corresponding to chemical polydispersity. We will point out, below, where the number of moments appearing in the excess free energy plays a role in the results we obtain.

\subsection{Chemically polydisperse polymer melt}

The diffusive flux for this system is specially simple. The mobility 
coefficients in eq \ref{eq:AB} reduce to
\begin{eqnarray}
A&=&-\frac{1}{\alpha N_p^2}\\
B&=&-\frac{1}{\alpha N_p}
\end{eqnarray}
With our assumptions, this is true for any melt when all polymers have the same length $N=N_p$.

We will focus on polymer melts whose excess free energy depends only on a finite set of moments. With chemical polydispersity, Flory-Huggins theory states that only the first moment of the chemical composition is involved~\cite{acp}:
\be
F^{ex}[{m_n({\bf r})}]=-\chi m_1^2 \label{eq:freeFH}
\ee
Correspondingly, the chemical potential gradients simplify to
\be
\nabla \mu_k=\frac{1}{\rho_k}\nabla\rho_k-2 \chi N_p \omega_{1k}\left[1+\frac{N_p}{24}\nabla^2\right]\nabla m_1
\label{eq:mugrad}
\ee
Note that in this case the function $W_0\equiv1+\frac{N_p}{24}\nabla^2$ does not depend on the polymer species. Since all polymers have the same length, we have (to second order in 
gradients),
\be
N_p\sum_k\omega_{nk}\rho_k({\bf r})=\left[1-\frac{N_p}{24}\nabla^2\right]m_n({\bf r})=W_0^{-1}m_n({\bf r})
\ee
where $W_0^{-1}$ is the inverse operator of $W_0$, likewise expanded to second order in gradients. 
Making use of this relation, the evolution equation for the species reads 
\begin{eqnarray}
\frac{d \rho_i}{dt} &=& -\frac{1}{\alpha N_p}\sum_k\nabla\cdot\left(-\frac{\delta_{ik}}{N_p}
+\rho_i\right)\rho_k\nabla\mu_k\nonumber\\
&=&-\frac{1}{\alpha N_p}\nabla\cdot\left\{-\frac{1}{N_p}\nabla\rho_i+2\chi (\omega_{1i}\rho_i
-W_0^{-1}m_1\rho_i)W_0\nabla m_1\right\}
\end{eqnarray}
where, due to incompressibility, $\nabla m_0=0$. This equation holds to leading nontrivial 
order in {\em separate} gradient expansions in the thermodynamics and in the relationships 
between fluxes and chemical potentials~\cite{footn}. This is the normal procedure in 
Cahn-Hilliard approaches to polymer dynamics and is the simplest theory able to predict 
a nontrivial wavevector dependence for the growth rates in spinodal 
decomposition\cite{olvera,clarke01,doi92}.

The evolution equation of the moment densities $m_i({\bf r})$ can in turn be written, 
within the same order of approximation, as
\be
\frac{d m_n}{dt } = D_0 \left\{\nabla^2 m_n+2  \chi N_p W_0\nabla\cdot((W_0^{-1} m_1)
 (W_0^{-1} m_n)-(W_0^{-1} m_{n+1})) W_0 \nabla m_1\right\}
\label{eq:mnFH}
\ee
with $D_0\equiv 1/(\alpha N_p^2)$. The diffusion coefficient 
$D_0$ sets the time scale, which  as expected, decreases proportionally to 
the increase of the friction coefficient.

This equation for the moment dynamics can be easily generalized to the case in which the 
excess free energy is a function of any finite set of moments. One gets in that case
\be
\frac{d m_n}{dt}=D_0 \left\{\nabla^2 m_n+N_p W_0\nabla\cdot\sum_{\alpha \beta}F^{ex''}_{\alpha\beta}\left((W_0^{-1} m_{\alpha+n})-(W_0^{-1} m_{\alpha})(W_0^{-1} m_n)\right) W_0 \nabla m_{\beta}\right\}
\label{eq:mnGen}
\ee
Note that for generic free energies this expression involves 
a linearization in moment gradients (which was not required for an excess free energy 
quadratic in $m_1$ as considered above). Apart from this, eqs \ref{eq:mnFH} and 
 \ref{eq:mnGen} have the same structure.  Each moment relaxes 
proportional to its own density deviation with a relaxation rate $D_0$, but is also coupled to 
all other species through the various moments that appear in the excess free energy of the model. 

Note also that the mobility does not play a major role; it only defines a characteristic time 
scale (which we will call the fast scale) $\tau_{f}(q) = 1/D_0q^2$ where $q$ is the wavenumber. 
In contrast, the coefficients  $F^{ex''}_{\alpha\beta}$ are related to the location of any 
phase transitions and/or spinodals. Proximity to these will control the form of any slow 
modes present. As a result, for an excess free energy with $n$ moments there are $n$ potential 
slow modes related to the thermodynamic forces of the system. This implies that, even if 
there is an hierarchy of kinetic timescales, it will be finite. 
(The situation will be quite different for length polydispersity, where there is no 
clear factorization of the moment evolution equations --- see section \ref{sect:length} below.)

For the chemically polydisperse melt there is only one moment appearing in the excess 
free energy ($m_1$).  The eigenfunctions of 
the diffusion matrix\footnote{ The diffusion matrix $D_{ij}$ is defined by writing 
eq \ref{eq:mnFH} as $\dot{m}_i(q)=D_{ij}(q)m_j(q)$ in Fourier space.} then have a 
very simple structure. In fact, all moments with $n>1$  are eigenvectors with the 
same eigenvalue $\lambda_n=-q^2 D_0$~\cite{footrates} with no wave vector dependence 
other than the usual diffusive one. On the other hand, when we linearize the 
evolution equation \ref{eq:mnFH} at wavenumber $q$ we find the one remaining eigenvalue
\be
\lambda_1 = -q^2 D_0 \left[1-2\chi N_p (\tilde{m}_2-\tilde{m}_1^2) W_0^2(q)\right] =
-D_0q^2\left[1-\frac{\chi}{\chi_{s}}(1-\frac{N_p}{12}q^2)\right]
\label{eq:eigenmelt}
\ee
where the tildes denote values in the homogeneous parent state about which the 
linearization is performed. We have also used that the interaction strength at the 
spinodal is given by $N_p\chi_{s}=1/(2 (\tilde{m}_2-\tilde{m}_1^2))$.
  As expected, $\lambda_1$ changes sign at the spinodal, and defines a slow mode in its 
vicinity (slow relative to all higher eigenmodes $\lambda_{n>1}$) that will control the 
early-stage kinetics of the phase separation process. During the initial stages of phase 
separation all moments $n>1$ will relax fast toward their corresponding local equilibrium 
values; but these values are controlled by $\lambda_1$ which evolves on a slower time scale. Note that this slow mode is not simply the first moment $m_1$, but a linear combination of 
all moments.

These results for the chemically polydisperse melt contrast strongly with those given in 
the previous literature for length polydispersity~\cite{warren,clarke01} as discussed further 
in section \ref{sect:length} below.

Specifically, since there is only one relevant moment, there is no distinction between a 
quenched and an annealed spinodal that can define different kinetic scenarios as described 
by Warren~\cite{warren} for length polydisperse solutions. Moreover, the slowest mode 
(which is the unstable one) is the only mode that involves the first 
moment. All moments of the distribution, 
except for the first one, will accommodate rapidly to the mean local composition, governed 
by the slow relaxation mode involving $m_1$. Finally, the wave vector dependence of the 
unstable mode is independent of 
polydispersity, and the same as in a monodisperse system. Only the overall amplitude 
(determined by the thermodynamics) is modified by 
polydispersity.

\subsection{Chemically polydisperse polymers in a passive solvent}

In this case, unlike the melt, the mobility plays a role beyond that of a simple scale factor 
in the dynamics. As explained previously, it is convenient to single out the solvent for 
special treatment en route to deriving dynamical equations for the moments $m_n$. 
For definiteness (and a clearer separation of timescales) we treat the case where the 
solvent has 
length $N_s \ll N_p$, although we continue to assume that all species are entangled. 
The ratio of the two lengths is defined 
as $r\equiv N_p/N_s$. 

The generic expression for the diffusive flux and the free energy will both 
depend now on the solvent number density. We first rewrite the relevant 
expressions in terms of polymer densities alone. These expressions will be 
valid both for this model and for length polydispersity addressed in 
section \ref{sect:length}. Due to incompressibility, the solvent number density 
is related to the monomer concentration ($m_0$) through
\be
\rho_s=\frac{1-m_0}{N_s}
\label{eq:incomp}
\ee
where we have disregarded the internal structure of the solvent in
writing this as a local relation. This is a good approximation at large
$r$ when the solvent is much smaller than the remaining polymers. In
what follows, we treat $r$ as a large parameter and are mainly concerned
with the leading behaviour in that limit; but where subleading
corrections are calculated, this is done without correcting
eq \ref{eq:incomp}.

We can now single out the solvent contribution both to the diffusive fluxes and the 
chemical potential gradients. For what we will call a `passive' solvent, its contribution 
to the free energy, eq \ref{eq:nonlocal} is simply to add a local entropic 
term $\rho_s(\ln\rho_s-1)$. Then
\be
\rho_s\nabla\mu_s = -\frac{1}{N_s}\nabla m_0
\ee
Note that for symmetric distributions in a passive solvent there is no tendency of the 
chemically polydisperse polymers to separate collectively from the solvent; the only 
tendency towards phase separation is within the polydisperse distribution itself. This 
ceases to apply for an asymmetric distribution of solute species. Note also that a more 
general solvent would introduce (as well as ignorable linear terms) terms in $m_0 m_1$ 
into the excess free energy, leading to somewhat more complicated phase 
behavior~\cite{acp} and dynamics.

Using the fact that $\rho_s$ depends only on the overall monomer concentration, the diffusive flux for the polymer species, given by eq \ref{eq:fluxpheno}, can now be expressed as a function of polymer variables only:
\begin{eqnarray}
{\bf J}_i&=&\sum_k'\left\{-\frac{\delta_{ik}}{\xi_i}+\frac{\phi_i}{\xi_i}+\frac{\phi_k}{\xi_k}-\left(\sum_j'\frac{\phi_j^2}{\xi_j}+\frac{(1-m_0)^2}{\xi_s}\right)\right\}\rho_k\rho_i\nabla\mu_k\nonumber\\
&-&\left[\frac{\phi_i}{\xi_i}+\frac{\phi_s}{\xi_s}-\sum_j'\frac{\phi_j^2}{\xi_j}-\frac{ (1-m_0)^2}{\xi_s}\right]\frac{\rho_i}{N_s}\nabla m_0
\end{eqnarray}
where the $'$ means that the sums run only over the solute polymer species.

For a solution of chemically polydisperse polymers, the excess free energy for a passive solvent may again be written within Flory Huggins theory as
\be
F^{ex}= -\chi m_1^2
\ee
where the interaction amplitude of interest (around the spinodal) obeys $\chi N_p \simeq 1$.
The chemical potential gradient is again given by eq \ref{eq:mugrad}. 
We can now derive expressions for the diffusive fluxes of each polymer species, 
following the same kind of derivation as for the melt. We take advantage of the 
fact that the friction coefficients for all species is the same 
(see eq \ref{eq:friction}) to arrive at
\begin{eqnarray}
{\bf J}_i&=&D_0\rho_i\sum_k'\left\{-\delta_{ik}+N_p \rho_k\right\}\nabla\mu_k\nonumber\\
&-&D_0\rho_i\left\{N_p(1-m_0)(r-1)\sum_k'\rho_k\nabla\mu_k+r(1+m_0(r-1))\nabla m_0\right\}
\label{eq:fluxi}
\end{eqnarray}
The first line coincides with the expression of the diffusive flux of a
species in the melt. The presence of a passive solvent has two effects
in the polymer dynamics, which correspond to the  two terms appearing in
the second line of eq \ref{eq:fluxi}. Gradients in the local monomer
concentration generate a flux of each species of polymers, but in
addition, the local presence of a solvent allows global polymer
rearrangements according to 
\be
\sum_k'\rho_k\nabla\mu_k=\frac{1}{N_p}W_0^{-1}\nabla m_0-2\chi (W_0^{-1}m_1)W_0\nabla m_1
\ee 
The analysis is very different from the melt case because now gradients in the local overall monomer concentration do not vanish.

Introducing the diffusive fluxes of the moment densities 
\be 
{\bf j}_n\equiv N_p \sum_k\omega_{nk}W_0{\bf J}_k
\ee
then using eq \ref{eq:fluxi} these read 
\begin{eqnarray}
{\bf j}_n&=&D_0\left\{-\nabla m_n+2\chi N_p W_0\left[(W_0^{-1}m_{n+1})-(W_0^{-1}m_{n})(W_0^{-1}m_{1})\right]W_0\nabla m_1\right.\nonumber\\
&+&2 \chi N_pW_0 (1-m_0)(r-1)(W_0^{-1}m_n)(W_0^{-1}m_1)W_0\nabla m_1\nonumber\\
&+&\left.W_0(W_0^{-1}m_n)\left[-r(1-m_0+m_0 r)+(2-m_0-r(1-m_0))W_0^{-1}\right]\nabla m_0\right\}
\end{eqnarray}
leading to
\begin{eqnarray}
{\bf j}_n&=&D_0\left\{-\nabla m_n+2\chi N_p \left[\tilde{m}_{n+1}-\tilde{m}_{n}\tilde{m}_{1}\right]W_0^2\nabla m_1\right.\nonumber\\
&+&2 \chi N_p (1-\tilde{m}_0)(r-1)\tilde{m}_n \tilde{m}_1 W_0^2\nabla m_1\nonumber\\
&+&\left.\tilde{m}_n\left[2-\tilde{m}_0-r (1-\tilde{m}_0)-r (1-\tilde{m}_0+\tilde{m}_0 r) W_0\right]\nabla m_0\right\}
\end{eqnarray}
where we have now linearized the fluxes, as appropriate for studying the structure factor during the early stages of phase separation. As in eq \ref{eq:eigenmelt}, the tilde refers to values of the moments evaluated at the reference state, instead of being dynamical quantities.

Again, the first line in the previous expression is the contribution that
we obtained for the melt. The second line requires the presence of a
solvent and corresponds to a correlated flow of all the species that
avoids segregation. Finally, the third line corresponds to the diffusive
flux of a given species induced by solvent gradients.

In terms of these moment fluxes, the evolution equation for the moments read
\be
\frac{d m_n}{dt}=-\nabla\cdot {\bf j}_n
\ee
which yields the final results for the linearized dynamic equations of
the moment densities themselves
\begin{eqnarray}
\frac{d m_n}{d t} &=& D_0 \left\{\nabla^2 m_n-2\chi N_p\left[\tilde{m}_{n+1}-\tilde{m}_n\tilde{m}_1 (1-(r-1)(1-\tilde{m}_0))\right]W_0^2\nabla^2m_1\right.\nonumber\\
&-&\left.\tilde{m}_n\left[ 1-(1-\tilde{m}_0)(r-1)-r(1+\tilde{m}_0(r-1))W_0\right]\nabla^2 m_0\right\}
\label{eq:solnmoms}
\end{eqnarray}

Although these equations are more complicated than for the case of a melt (eq \ref{eq:mnFH}), they do retain a relatively simple 
quasi-triangular form~\cite{footquasitri} which allows us to analyze in some detail the 
structure of the eigenvalues. The fact that all moments only couple to the first two 
moments ($m_0$ and $m_1$) imply that, except for these two, the rest are 
degenerate eigenfunctions. Hence we may set $\lambda_n=-D_0 q^2$ for $n =2,3,...\,$. The interesting dynamical behavior lies then in the remaining two eigenmodes, which are 
the only ones that involve $m_0$ and $m_1$.

\subsubsection{Dependence on solvent size ratio}
In general the expressions for the two eigenvalues is quite involved,
but given our assumptions so far, they can be found for arbitrary
$r$. However, the derivation was subject to eq \ref{eq:incomp} which
itself is valid for large $r$ only. Hence we relegate the general
expressions to Appendix \ref{app:1}. 

In the limiting case of very large polymers (i.e. $r\gg 1$) the two
 eigenvalues simplify to
\begin{eqnarray}
\lambda_0&=&-q^2 D_0 r^2 \tilde{m}_0^2 W_0(q)\\
\lambda_1&=& -q^2 D_0\left[1-\frac{\chi}{\chi_s}W_0(q)^2\right]
\label{eq:eigenasym}
\end{eqnarray}
where $\chi_s=\tilde{m}_0/(2 (\tilde{m}_0 \tilde{m}_2-\tilde{m}_1^2))$  is the 
spinodal value of the interaction parameter (to leading order in large $r$). 
For the particular case of a symmetric parent (i.e., a symmetric
distribution of the polydisperse species in the homogeneous phase prior
to the quench) the expressions for the the eigenvalues are particularly
simple. In this case $\tilde{m}_1=0$ in the initial state and the eigenmodes that 
involve $m_0$ and $m_1$ do not mix them; their eigenvalues (exact
to order $q^4$) are
\begin{eqnarray}
\lambda_0&=&-D_0q^2\left[1-\tilde{m}_0\left\{1-(1-\tilde{m}_0)(r-1)-r [1+\tilde{m}_0 (r-1)]W_0(q)\right\}\right]\\
\lambda_1&=&-D_0 q^2 \left(1-2 \chi \tilde{m}_2 W_0(q)^2\right)=- D_0 q^2
\left(1-\frac{\chi}{\chi_s} W_0(q)^2\right)
\label{eq:eigensym} 
\end{eqnarray}
with $\chi_s=1/(2 \tilde{m}_2)$ the spinodal value of the interaction parameter for a symmetric parent. Note that eqs \ref{eq:eigenasym} and \ref{eq:eigensym} have the 
same form as the relevant eigenvalue in the melt case,
eq \ref{eq:eigenmelt}. The only difference is the location of the spinodal; it
is through the thermodynamic sector only that the solvent, and likewise the polydispersity, can enter. This shows that in the limit of high size ratio ($r\gg1$) the main features of mode structure are, within our model, universal.

Likewise if we look at the wave vector dependence of the eigenvalues $\lambda_0$ and $\lambda_1$, we see that it does not depend directly on 
polydispersity. This behavior is again analogous to that of the melt, and can be 
attributed to the absence of coupling of the unstable mode to all the rest of 
the modes. This is contrary to what happens for length polydispersity, as we discuss in section \ref{sect:length}.

For both symmetric and asymmetric parents, in the limit $r \gg 1$, the two eigenvalues scale asymptotically as
\begin{eqnarray}
\lambda_0&\sim&-\frac{q^2 \tilde{m}_0^2}{\alpha N_s^2}\\
\lambda_1&\sim&-\frac{q^2}{\alpha N_p^2}\left[1-\frac{\chi}{\chi_s}\right]
\end{eqnarray}
but the convergence to this limit as $r$ is increased strongly depends
on parental asymmetry. In Figure1 we show the wave vector dependence of
the unstable eigenmode, $\lambda_1$ for an exponential parent
($\rho_n\sim\exp(a f_n)$) as a function of the size ratio $r$ for
different values of $a$. When the asymmetry parameter $a$ is increased,
the wavevector dependence becomes more sensitive to the size ratio $r$
(in fact, for $a=0$ there is no dependence on $r$;
cf. eq \ref{eq:eigensym}). The dependence on size ratio becomes also
more relevant as the polymer dilution increases. For high asymmetry and
dilution (Figure1c), the convergence to infinite aspect ratio becomes much
slower and it is not monotonous for intermediate values of $r$. In
Figure2 we show the wave vector dependence of the stable mode, $\lambda_0$, for two
 disparate values of the asymmetry at high dilution. The dependence on
$q$ is monotonous as anticipated. In all cases the curves converge in the
limit of large aspect ratios (note that because of the different
scaling of the two eigenvalues $\lambda_0$ and $\lambda_1$ we have
scaled here  $\lambda_0$ by $r^2$). For large asymmetry (Figure2b),
also for this stable mode the convergence to the high aspect ratio limit becomes
slower and it is non-monotonous for intermediate $r$ values.

\subsubsection{Qualitative behaviour at large size ratio}
The above analysis shows that, for large size ratios between solute and 
solvent ($r\gg 1$), $\lambda_0$ is 
much larger than the rest of the eigenvalues. 
This means that the mode connected with 
$\lambda_0$ (which, for a symmetric parent only, does not involve $m_1$)
 will grow or decay much faster than all the rest
of the modes, whose rates scale with the inverse of the solute polymer
length $N_p$, defining for $n>1$ a common time scale $\tau_{f}(q) =
1/D_0q^2$. The latter was already identified as the fast time
scale in the melt case. The corresponding modes remain fast relative not
to the solvent mode (which is faster), but to a single slow mode. This
is governed by $\lambda_1$, which splits off from this family of modes
with a time scale that diverges on the spinodal. 

This means that during the early stages of phase separation (or during the 
decay of an external perturbation within the stable region of the phase 
diagram), the overall monomer density is first equilibrated (if necessary); 
on this timescale the remaining moments are effectively quenched. In a 
second stage the chemical 
composition will relax towards its equilibrium distribution (with the
monomer density slaved to this). Close to the spinodal this chemical
relaxation can, just as in the melt case, be separated into two
stages. These entail relatively rapid equilibration of the higher
moments to a state of local equilibrium set by the slow evolution of the
mean composition variable $m_1$ (which is the local excess of A over B
monomers in the case of a random copolymer system). It is notable that,
for an asymmetric parent, the two relevant modes include nontrivial linear 
combination of  $m_0$ and $m_1$, while for a symmetric parent these two 
moments appear separately in the corresponding eigenfunctions. This is 
because of the thermodynamic
coupling which, for an asymmetric parent, requires chemical separation
to be accompanied by changes in the overall monomer density.

This picture is somewhat analogous to one described by Warren for the case 
of length polydisperse polymers undergoing phase separation from an 
incompatible solvent~\cite{warren}. Warren developed a picture in 
which there were two separate spinodal curves: the equilibrium one and, 
within it, a quenched one to describe the dynamics of a hypothetical 
system whose overall monomer density could change but where the relative 
prevalence of different chain lengths could not change locally. Between 
these two spinodals, the kinetics were argued to be controlled by the 
slow sorting out of the polydisperse species. 

For chemically polydisperse chains in a solvent, a similar scenario can
arise. However, for the passive solvent considered here, this can only
happen for an asymmetric parent since otherwise the phase separation is
towards two states of equal solvent density; this density coincides with
that in the initial state. Accordingly there is no driving force for
phase separation in a symmetric system on the time scale of the fast
(solvent) mode as may be confirmed by constructing the relevant free
energy with quenched $\langle f\rangle = \tilde{m}_1/\tilde{m}_0$.  If
we consider a fully quenched situation where all higher moments are slave to
 $m_0$, ( i.e. higher moments evolve as $m_n=\langle f^n\rangle m_0$),
 its linearized evolution equation is
\begin{eqnarray}
\frac{d m_0}{dt}&=&D_0\left\{1-2\chi\tilde{m}_0\langle
f\rangle^2(1-\tilde{m}_0)\left[1+\tilde{m}_0(r-1)\right]W_0^2\right.\nonumber\\
&+&\left.\tilde{m}_0\left[-1+(1-\tilde{m}_0)
(r-1)+r (1+(r-1)\tilde{m}_0)W_0\right]\right\}\nabla^2 m_0
\end{eqnarray}
which for large $r$ will give a relaxation rate scaling with the fastest mode of 
the system. This leads to a rapid growth of fluctuations, but only inside a region 
where the quenched system is unstable; such a region is present only for an asymmetric parent. 

In Figure3 we show the quenched and annealed spinodals for a chemically
polydisperse mixture with an exponential parent shape ($\rho_n\sim \exp(a
 f_n))$ for different values of the parameter $a$ (which controls the degree
of asymmetry of the mixture) and of the size ratio $r$. One can see that in the 
regime where polymers are much larger than the solvent, the two spinodals are 
far apart. There is thus a broad region, in between the two spinodals, where 
the density relaxation will take
place only through a slow sorting of the chemical species. The slowest
process will then control the initial steps of the phase separation
process, determining the relevant time scale.

\section{Length polydispersity}
\label{sect:length}

\subsection{Structure of equations}
Before analysing the effect of length polydispersity in detail, it is instructive to 
compare the structure of eq \ref{eq:mik} with the analogous one derived by 
Clarke~\cite{clarke01}. He obtained diffusive fluxes for the monomer concentrations, 
disregarding (as we do) the internal structure of the chains when constructing the relation between fluxes and chemical potential gradients. (The internal structure enters only when calculating these gradients.) We may rewrite our results in terms of the {\em monomeric} flux of each species $i$ (that 
we will call $\tilde{{\bf J}}_{i}$) starting from  eq \ref{eq:mik}. We get,
\be
\tilde{{\bf J}}_{i} = \sum_k\left\{-\frac{\delta_{ik}}{\xi_i}+\left(\frac{\phi_i}{\xi_i}
+\frac{\phi_k}{\xi_k}\right)-\sum_j\frac{\phi_j^2}{\xi_j}\right\}\phi_i\phi_k\nabla\tilde{\mu}_{ k}\equiv \sum_j\Lambda_{ij}\nabla\tilde{\mu}_j
\label{eq:mikphi}
\ee
where $\tilde{\mu}_{k}$ is the monomeric chemical potential of species $k$, i.e. $\tilde{\mu}_{ k} = \delta F/\delta \phi_k$. The mobility matrix $\Lambda_{ij}$ coincides with the one derived by Clarke (cf. eq 22 of Ref.~\cite{clarke01}) if we identified the phenomenological coefficients $\tilde{\lambda}_i$~\cite{notation} introduced in that paper as
\be
\tilde{\lambda}_i\equiv\frac{\phi_i^2}{\xi_i}
\ee
The basic differences between Clarke's approach and ours is then that while
 he considered diffusive fluxes referred to a common 
velocity fixed  through incompressibility, in our case the diffusive
fluxes are defined with respect to 
the barycentric velocity. In this way we get diffusive fluxes that are 
strictly proportional to the chemical potential gradients of each species.

Nonetheless this comparison is suggestive, because the phenomenological 
model of Clarke is claimed to apply both to entangled and unentangled mixtures.
In the former case he  suggests that $\tilde{\lambda}_i$ should scale as the inverse 
of the polymer length, $\tilde{\lambda}_i\sim 1/N_i$. We have derived this result 
in section \ref{sec:micro}.
Brochard's theory (which was our own starting point in section \ref{sec:micro}), and 
also Kramer's theory~\cite{Kramer} (on which Clarke based some of his reasoning) both 
assume that the diffusion is controlled by the fastest species (they are accordingly 
referred to as `fast mode theories')~\cite{footfast}. Hence, it is not surprising that 
we get mobility coefficients that concur with those posited by Clarke on the basis 
of Kramer's model.

For the unentangled case, Clarke argues that, because of the different scaling of the 
self-diffusion coefficient with polymerization (Rouse model~\cite{DoiEdwards}) one 
can assume that in this situation $\tilde{\lambda}_i$ is a constant instead.  
Although we have derived the 
expressions for the diffusive fluxes within the context of the tube model, 
if we assume, following Clarke, that their overall functional form is the same in 
the unentangled case we can use our results in both situations simply by changing 
the dependence of the friction coefficients on the polymer lengths. For chemical 
polydispersity as discussed above, this means modifying only one or two global 
parameters (solvent and solute chain mobilities) of the theory; for length polydispersity 
the effect is more complicated and discussed in section
\ref{subsec:unent} below.

\subsection{Length-polydisperse chains in a solvent}
For length-polydisperse chains in a solvent, treated within Flory Huggins theory, the 
free energy is written as 
\be
{\cal F} =\int d{\bf r} \sum_k\rho_k({\bf r})(\ln \rho_k({\bf r})-1)+\chi m_0({\bf r})(1-m_0({\bf r}))
\ee
where we have initially chosen the sum over species to {\em include} the solvent chains, 
while the moments are the linear nonlocal combinations of the polymer number densities 
defined in eq \ref{eq:omegamoments} with $\omega_{nk} = N_k^n$, and the solvent 
chains {\em excluded}. Note that, as described following eq \ref{eq:omegamoments}, $m_0$ 
is the total volume fraction occupied by solute monomers. The chemical potentials of the 
species are accordingly
\begin{eqnarray}
\mu_k&=&\log\rho_k+\chi N_k-2\chi N_k\int \tilde{w}_k({\bf r-r'})m_0({\bf r'})\, d{\bf r'}\\
\mu_s&=& \log\rho_s=\log\left(\frac{1-m_0}{N_s}\right)
\end{eqnarray}
where we again neglect the spatial extent of the solvent chains, presuming them 
small compared to the typical size of the solute polymers. 

As we did for chemical polydispersity, we now choose to treat the
solvent as a separate species and derive equations for the diffusive
fluxes of the solute polymers with the solvent coupling eliminated
(statically and dynamically) via the incompressibility constraint,
eq \ref{eq:incomp}. Using this relation, and the fact that
$W_{0i}-1$ is linear in species length~\cite{rhotomfoot}, the diffusive flux of species $i$ 
can now be found to second order in gradients. The full result is given in 
Appendix \ref{app:2} and is quite complicated. Truncating further to first order in 
gradients, the fluxes obey
(with $\alpha = \xi_0/N_e$)
\begin{eqnarray}
\alpha {\bf J}_i &=& -\frac{1}{N_i^2}\nabla\rho_i+\rho_i \nabla m_{-2}
+\left[\frac{\rho_i}{N_i}-\rho_i\left(m_{-1}+\frac{1-m_0}{N_s}\right)\right]\nabla
 m_{-1}\nonumber\\
&+&\rho_i\left[2\chi(1-m_0)-\frac{1}{N_s}\right]\left(\frac{1}{N_i}-m_{-1}
+\frac{m_0}{N_s}\right)\nabla m_0
\end{eqnarray}

The completely different appearance of this expression from the one obtained 
previously for \\chemical polydispersity arises because the friction factor $\xi$ 
in eq \ref{eq:mik} depends explicitly on \\polymerization index $N_i$ which is 
now the polydisperse variable and not the same for all solute chains. Because of 
the factor $N_k$ arising in the definition of the moments, eq \ref{eq:omegamoments}, 
the overall number density of solute chains is $m_{-1}$.
Since in addition the friction appears as a denominator in eq\ref{eq:mik}, two 
negative moments appear in the above equation.

Rewriting this in terms of diffusive fluxes ${\bf j}_n$ for the moment densities
  $m_n({\bf r})$, we obtain in the linearized regime
\begin{eqnarray}
\alpha{\bf j}_n &=& -\nabla m_{n-2}+\tilde{m}_n\nabla m_{-2}+\left[\tilde{m}_{n-1}-\tilde{m}_n \tilde{m}_{-1}-\tilde{m}_n \frac{1-\tilde{m}_0}{N_s}\right]\nabla m_{-1} \nonumber\\
&+&\left[2\chi(1-\tilde{m}_0)-\frac{1}{N_s}\right]
\left(\tilde{m}_{n-1}-\tilde{m}_n\tilde{m}_{-1}+\frac{\tilde{m}_0\tilde{m}_n}{N_s}\right)\nabla m_0
\label{eq:jnlength}
\end{eqnarray}
In Appendix \ref{app:2} we provide the corresponding expression to second
order in gradients, which can be used, in principle, to study the wavevector dependence of the relaxation modes.

But even in the long wavelength limit, the set of equations is no longer at all tractable: moment $m_n$ is coupled to $m_{n-2}$, and all moments are coupled to $m_0$, $m_{-1}$ and $m_{-2}$. The higher moments are no longer simple eigenfunctions. Also, there is the additional complication that moments of negative powers of the 
chain length have entered the description. 
Since we know that the set of functions $\omega_i(N)=N^i$ 
form a basis for $i=0,1,...$, we can in fact express each negative moment as a linear 
combination involving only moments with positive index. In doing so, though, we  reach an expression for the 
diffusive fluxes of positive moments which couple to gradients of {\em all} 
the other positive moments. It is clear that the diffusion matrix is no longer 
truncatable and an infinite hierarchy of dynamical modes can be
expected. Note that this hierarchy remains compatible with a gross
separation into two basic timescales when solvent and solute chains are
very different in length. One can still expect a single mode scaling
with the inverse solvent size, with the remainder involving the solute
chain sizes instead; the hierarchy controls the details of how the
solute chains segregate by size.

\subsection{Approximate truncation of the dynamics}

Following Warren~\cite{warren}, we can try to analyze in more detail the structure of the dynamic modes for length polydispersity by considering the dynamical equations of a finite 
subset of moments created by simply truncating the diffusion matrix. Thus we 
assume that the flux of any species is only coupled to spatial gradients 
of a finite set of moments. {\em A priori}, there is no guarantee that such a 
process will preserve the character of the modes; we will attempt to verify, {\em a posteriori}, that this procedure is in fact sensible.

Let us consider a two-moment description and assume that the two relevant modes are the volume fraction, $m_0$, and 
the chain number density, $m_{-1}$. According to eq \ref{eq:jnlength}, their 
fluxes are coupled to gradients of other moments. However, the assumption 
that only $m_0$ and $m_{-1}$ are relevant allows us to express other moments 
by the corresponding projection into the subspace spanned by the retained 
moments. To be precise, the weight functions $\omega_k$ with $k$
different from 0 and $-1$ will be approximated as a linear combination
of those two weight functions, using the shape of the parent. In this way, we can write
\be
\omega_{-k}=\frac{\langle N^{-k}\rangle\langle N^{-2}\rangle -\langle N^{-k-1}\rangle\langle
N^{-1}\rangle}{\langle N^{-2}\rangle-\langle
N^{-1}\rangle^2}\omega_{0}+\frac{\langle N^{-k-1}\rangle-\langle N^{-k}\rangle\langle N^{-1}\rangle}{\langle N^{-2}\rangle-\langle
N^{-1}\rangle^2}\omega_{-1}
\ee
where $\langle ... \rangle$ means an average over the chain length
distribution in the parent.  With this choice we ensure
that  the average of $\omega_{-k}$ and $\omega_{-k}N^{-1}$ over the
parent remains correct.

We can now approximate the two additional moments ($m_{-2}$ and $m_{-3}$) appearing in the
evolution equations of the monomer and chain number densities as linear
combination of the retained quantities. In this way, we disregard the coupling
of all moments except for $m_0$ and $m_{-1}$, getting a closed set of
equations. 
The required projection of moments $m_{-3}$ and $m_{-2}$ read
\begin{eqnarray}
m_{-3}&=&\frac{\langle N^{-3}\rangle\langle N^{-2}\rangle-\langle N^{-4}\rangle\langle N^{-1}\rangle}{\langle N^{-2}\rangle-\langle
N^{-1}\rangle^2}m_{0}+\frac{\langle N^{-4}\rangle-\langle N^{-3}\rangle\langle N^{-1}\rangle}{\langle N^{-2}\rangle-\langle
N^{-1}\rangle^2}m_{-1} \equiv \beta_{\bar{3}\bar{1}}m_{-1}+
\beta_{\bar{3}0}m_{0} \nonumber\\
m_{-2}&=&\frac{\langle N^{-2}\rangle^2-\langle N^{-3}\rangle\langle N^{-1}\rangle}{\langle N^{-2}\rangle-\langle
N^{-1}\rangle^2}m_{0}+\frac{\langle N^{-3}\rangle-\langle N^{-2}\rangle\langle N^{-1}\rangle}{\langle N^{-2}\rangle-\langle
N^{-1}\rangle^2}m_{-1} \equiv \beta_{\bar{2}\bar{1}}m_{-1}+
\beta_{\bar{2}0}m_{0}
\label{eq:projection}
\end{eqnarray}
The averages $\langle\dots\rangle$can be expressed in terms of the moments of the parent (which are constants, not dynamical variables) through $\langle N^n\rangle = \tilde{m}_{n-1}/\tilde{m}_{-1}$.

The diffusive fluxes for the retained moments then reduce to
\begin{eqnarray}
\alpha {\bf j}_{-1} &=& \left[-\beta_{\bar{3}\bar{1}}+\beta_{\bar{2}\bar{1}}m_{-1}+m_{-2}-m_{-1}^2-(1-m_0)\frac{m_{-1}}{N_s}\right]\nabla m_{-1}\nonumber\\
&+& \left\{-\beta_{\bar{3}0}+\beta_{\bar{2}0}m_{-1}+2\chi (1-m_0)\left[m_{-2}-m_{-1}^2+\frac{m_{-1}m_0}{N_s}\right]\right.\\ &+& \left.\frac{1}{N_s}\left(m_{-1}^2-m_{-2}-\frac{m_{-1}m_0}{N_s}\right)\right\}\nabla m_0\\
\alpha {\bf j}_0 &=& (1-m_0)\left[-\beta_{\bar{2}\bar{1}}+m_{-1}-\frac{m_0}{N_s}\right]\nabla m_{-1}\nonumber\\
&+&(1-m_0)\left\{-\beta_{\bar{2}0}+2\chi \left[m_{-1}(1-m_0)+\frac{m_0^2}{N_s}\right]-\frac{m_{-1}}{N_s}-\frac{m_0^2}{(1-m_0) N_s^2}\right\}\nabla m_0
\end{eqnarray}
In the linearized regime, each of the factors preceding the gradient operator can be evaluated using the parent, in which 
$\tilde{m}_k=\tilde{m}_0 \frac{\langle N^{k+1}\rangle}{\langle N\rangle}$.
Using also the size ratio $r=\langle N\rangle/N_s$, the linearized equations read then
\begin{eqnarray}
\alpha{\bf j}_{-1}&=&\left[-\langle
N\rangle^2\beta_{\bar{3}\bar{1}}+\langle N\rangle\beta_{\bar{2}\bar{1}}
\tilde{m}_0+\tilde{m}_0(\langle N^{-1}\rangle\langle N\rangle-\tilde{m}_0-r(1-\tilde{m}_0))\right]\frac{1}{\langle N\rangle^2}\nabla
m_{-1} \nonumber\\
&+&\left\{-\langle
N\rangle^3\beta_{\bar{3}\bar{0}}+\langle
N\rangle^2\beta_{\bar{2}\bar{0}}\tilde{m}_0\right.\nonumber\\
&-&\left. 2\chi\langle N\rangle(1-\tilde{m}_0) \tilde{m}_0 \left[\langle N^{-1}\rangle \langle N\rangle
+\tilde{m}_0 (r-1)\right]-r \tilde{m}_0 (\tilde{m}_0 (r-1)+\langle N^{-1}\rangle \langle N\rangle)\right\}\frac{\nabla m_0}{\langle N\rangle^3}\nonumber\\
\alpha{\bf j}_{0}&=&(1-\tilde{m}_0)\left[-\langle
N\rangle\beta_{\bar{2}\bar{1}}+(1-r)\tilde{m}_0\right]\frac{1}{ \langle
N\rangle}\nabla m_{-1}\nonumber\\
&+&(1-\tilde{m}_0)\left[-\langle N\rangle^2\beta_{\bar{2}\bar{0}}+2\chi\langle
N\rangle \tilde{m}_0 (1+\tilde{m}_0(r-1))-\tilde{m}_0 r\left(1+\frac{\tilde{m}_0 r}{1-\tilde{m}_0}\right) \right]\frac{1}{\langle N\rangle^2}\nabla m_0
\end{eqnarray}

These expressions remain fairly cumbersome. To proceed further and gain
explicit expressions for the eigenvalues, we will
consider only one specific example, where the parent has a uniform distribution with mean length $\langle N\rangle$ and half width 
$\Delta$. We will work with the preceding expressions which are lowest order in the gradient expansion. To simplify things further we will consider the limit of a narrow parent ($\langle N\rangle \gg \Delta$) and large size ratio ($r \gg 1$). To lowest order, the two eigenvalues then simplify to
\begin{eqnarray}
\lambda_0&=&-D_m q^2\left\{ (1+\tilde{m}_0 (r-1))^2
\left(1-\frac{\chi}{\chi_s}\right)+\frac{(1-\tilde{m}_0)(1+\tilde{m}_0 r)
 \left[1+\left(1-\frac{\chi}{\chi_s}\right)(3+2 \tilde{m}_0 r)\right]}{3
 \left[1-\left(1-\frac{\chi}{\chi_s}\right) (1+ \tilde{m}_0r)^2\right]} 
\left(\frac{\Delta}{2\langle N\rangle}\right)^2\right\} \nonumber\\
&\simeq&- \frac{q^2\tilde{m}_0^2}{N_s^2}
 \left(1-\frac{\chi}{\chi_s}\right) -\frac{2 q^2}{3\langle N\rangle^2}(1-\tilde{m}_0)\left(\frac{\Delta}{2 
\langle N\rangle}\right)^2 \nonumber\\
\lambda_1&=&-D_mq^2\left\{ 1+ \frac{63+5 \tilde{m}_0 r-5 \tilde{m}_0^2 r -
 \left(1-\frac{\chi}{\chi_s}\right) (1+\tilde{m}_0 r) (43+48 \tilde{m}_0 r)}{15
 \left[1-\left(1-\frac{\chi}{\chi_s}\right) (1+\tilde{m}_0r)^2 \right]}\left( 
\frac{\Delta}{2\langle N\rangle}\right)^2 \right\}\nonumber\\
&\simeq& -\frac{q^2}{\langle N\rangle^2} 
\left[1+\frac{16}{5}\left(\frac{\Delta}{2\langle N\rangle}\right)^2\right]
\label{eqn:prev}
\end{eqnarray}
where $D_m\equiv 1/\alpha\langle N\rangle^2$ is a characteristic diffusion
 coefficient for the polymer solution, and where the value of the spinodal strength $\chi_s$ coincides with its
 monodisperse counterpart to the order of validity of the previous expressions.
  
We see that (at least in the low $q$ limit studied here) the two modes show a structure analogous to, but crucially different from,
that described before for chemical polydispersity. The generic tendency at large size ratio $r$ is for $\lambda_0$ to greatly exceed
$\lambda_1$, with the former defining a fast time scale and the latter a slow one. But this tendency is reversed close enough to the spinodal, where $\lambda_0$ now vanishes (not $\lambda_1$ as in the chemically polydisperse case). The difference arises because for chemical polydispersity the excess free energy involves the first moment of the polymer species, while here it involves the monomer concentration (zeroth moment). The nominally `fast' mode is the thermodynamically unstable one; this was anticipated in Warren's original proposal~\cite{warren}. Between the quenched and the annealed spinodals, the phase
separation can only proceed initially via species sorting, whereas beyond the quenched spinodal, the fast mode ($\lambda_0$) will drive the phase separation process from the outset.

 In order to see if the assumption that only two moments matter is too 
crude, we have repeated the analysis assuming that the relevant moments are the 
three to which all the other moments are coupled,  i.e.  $m_{-2},m_{-1},m_0$. For
the linearized regime we give the expressions for the corresponding
fluxes in Appendix \ref{app:3}. In the limit of a narrow parent and large asymmetry the qualitative behavior of the three eigenmodes, found numerically, is the same as the one discussed with the previous approximation
,  in the sense that in both cases one eigenvalue
($\lambda_0$) is generically larger than the rest (since it scales with
the inverse solvent size), but crosses over to become smaller than the
others in a narrow region as the spinodal is reached\cite{perturbation}.

\subsection{Unentangled solution of length-polydisperse chains}
\label{subsec:unent}
As we have discussed previously, if we follow Clarke~\cite{clarke01}, we can assume that the expressions for the diffusive fluxes derived for an entangled mixture in eq \ref{eq:mik} keep their general form for an unentangled solution. The only difference is that in this case the friction coefficients are proportional solely to the monomer concentrations, $\xi_i\equiv \alpha \phi_i$, and beyond this have no explicit dependence on polymer length.
In this case the diffusive fluxes of species, eq \ref{eq:mik} reduce to
\be
{\bf J}_i = \frac{1}{\alpha}\left\{\sum_k'\left[-\frac{\delta_{ik}}{N_i}+\rho_k\right]\rho_i\nabla\mu_k+\rho_i\rho_s\nabla\mu_s\right\}
\ee
and those of the moment densities become
\begin{eqnarray}
{\bf j}_n&=&\frac{1}{\alpha}\left\{-\nabla m_{n-1}+m_n\nabla
m_{-1} \right.\nonumber\\
&+&\left.\left[2\chi \left[m_n(1-m_0)+\frac{1}{24}(m_{n+1}\nabla^2+m_n\nabla^2m_1-m_nm_1\nabla^2)\right]-\frac{m_n}{N_s}-\frac{1}{24} m_n\nabla^2\right]\nabla m_0\right\}
\end{eqnarray}
where we have related local moments of chain number
densities to local moments of monomer \\concentrations~\cite{rhotomfoot}.

To analyze the onset of phase separation, we can linearize the previous
expression to give
\be
{\bf j}_n=\frac{1}{\alpha}\left\{-\nabla m_{n-1}+\tilde{m}_n\nabla
m_{-1}+\tilde{m}_n\left[2\chi\langle N\rangle (1-\tilde{m}_0)-r\right]\frac{\nabla m_0}{\langle N\rangle}\right\}
\ee
If we assume that we can isolate the dynamics of $m_0$ and $m_{-1}$ by
closing their dynamic equations as done in the previous section, we
obtain for a parent with narrow uniform distribution, in the limit of a narrow distribution (i.e. $\Delta \langle 1$) 
\begin{eqnarray}
\lambda_0&=&-\hat{D}_m q^2\left[(1+\tilde{m}_0 (r-1))\left(1-\frac{\chi}{\chi_s}\right)+\frac{(1-\tilde{m}_0)\Delta^2}{3 \left[-1+\left(1-\frac{\chi}{\chi_s}\right) (1+ \tilde{m}_0(r-1))\right]} \right]\nonumber\\
\lambda_1&=&-\hat{D}_m q^2\left\{1+\Delta^2\left[\frac{22}{15}-\frac{1-\tilde{m}_0}{15\left[-1+\left(1-\frac{\chi}{\chi_s}\right) (1+ \tilde{m}_0(r-1))\right]}\right]\right\}
\end{eqnarray}
where, as before $\hat{D}_m\equiv 1/\alpha\langle N\rangle$ is a
characteristic diffusion coefficient for the polymer solution. Again, we have kept only the
dominant contribution in $r$. The value of $\chi_s$ is the same as
for the entangled solution, and deviations from its monodisperse value
are also negligible in the narrow limit examined.

The eigenmodes here are thus essentially equivalent to the situation
described above for the entangled solution. We again observe the appearance of a `fast' mode which nonetheless becomes slow very close to (and then unstable at) the spinodal. Away from there, the generic ratio between the two
eigenmodes in this case is not as big as for the entangled case, because the
dependence of the friction coefficients on polymerization is not as
strong here.

Incidentally, note that in the hypothetical case that the friction
coefficient would have been proportional to the chain number density,
then all moments would have coupled to themselves and to $m_0$, $m_1$
and $m_2$. In this case, the only nontrivial eigenmodes would have been the three involving these three moments. A simple picture in terms
of the moments is hence recovered exactly in that case.

\section{Conclusions}
\label{sec:conclusions}

In this paper we have analyzed the dynamics of polydisperse polymeric
materials. Polydispersity affects both the thermodynamics and the
mobilities of the system. For a polymeric dense solution, for any
polydispersity other than length polydispersity, the mobilities are the
same for all species. In such case the moment structure of the excess
free energy carries over to the dynamics of the system, and it is
possible to  obtain a simple picture in terms of the moments that appear
in the excess free energy. Those are the relevant ones in the linearized
dynamic regime, and we have analyzed their implications for the
particular case of chemical polydispersity.  

The moment structure of the equations of motion makes it possible to map
the polydisperse system into an effective binary mixture with a
judicious choice of ``effective'' species. These fictitious species are
linear combinations of chain densities such as the monomer concentration
and the mean chemical composition. Previous work has not exploited this
choice but instead used a small number of discrete values of the
polydisperse variable to impersonate the case of continuous polydispersity~\cite{olvera2}.

We have seen that this finite moment structure implies in turn that
there exists a finite hierarchy of relevant modes in the system. These
modes define a fast and a slow time scale that will affect how kinetics
takes place depending on the position in the phase diagram. Hence, there will
be situations where the fast  separation of the overall density will
be dominant at short times (and only later the local chemical
composition will rearrange on a slower time scale), but in certain instances it will be the
chemical sorting of species which will control the kinetics from the
beginning.

The difficulties in the treatment of length polydispersity come from the
interplay of dynamics and thermodynamics. In the approach of Warren~\cite{warren}
such a coupling is not accounted for. Were no such coupling present, our
analysis could follow that for chemical polydispersity; we might even be
tempted to average (over the parent) the diffusion matrix and the chemical
potential gradients separately. However, our analysis suggests that this
constitutes a poor description of the true dynamics.

The dependence of mobility on polymer length makes the treatment of
length polydispersity more complicated due to the interplay between free
energy changes and mobility couplings. Starting from a  fast mode
theory, we have derived the equations of motions for the moments. By keeping both positive and negative powers of the length, we arrived at
a set of dynamical equations that, while more complicated than for
chemical polydispersity, retain a relatively simple structure. In fact, the evolution of a given moment is only coupled to
gradients of three other moments (whereas in terms of the species, all species
are coupled to gradients of all the other species).  The fact that we
have to keep negative powers of $N_i$ stems from the fact that we are
dealing with a fast mode theory. If we re-express those powers in terms of positive powers of $N_i$, then
the dynamics of the remaining moments couple to gradients of all the
others. This shows that a judicious choice of moments (including negative
ones where appropriate) is central to the simplification of the dynamics.

The study of the linear dynamics shows also the appearance of a fast a a
slow time scales (as was the case for chemical polydispersity),although
in this case the role of the fast and slow modes are inverted. A
detailed analysis of the modes deserves further work to elucidate the
structure of the slow time scale.

\section{Acknowledgements}
We acknowledge important discussions with N. Clarke and R.M.L. Evans during the preliminary stages of this work. We also thank P. Sollich and P.B. Warren for illuminating discussions. Work funded in part by EPSRC GR/M29696.

\appendix
\section{Eigenvalues for chemical polydispersity}
\label{app:1}

For a chemically polydisperse solution in the presence of a passive
solvent, the two relevant eigenvalues in the linearized regime are
obtained from eq \ref{eq:solnmoms}. The only coupled moments are $m_0$
and $m_1$. Therefore, the  relevant linear system of equations is then defined by the matrix,

\footnotesize
\begin{equation}
\left(\begin{array}{l l}
	-q^2D_0\left\{1-\tilde{m}_0\left[1-(1-\tilde{m}_0)(r-1)-r \left[1+\tilde{m}_0(r-1)\right]W_0(q)\right]\right\}&2\chi D_0 N_p\tilde{m}_1
	(1-\tilde{m}_0)\left\{1+\tilde{m}_0 (r-1)\right\}W_0(q)^2 q^2 \\
 & \\
 & \\
 & \\
	q^2 D_0 \tilde{m}_1 \left\{1-(1-\tilde{m}_0) (r-1)-r \left[1+\tilde{m}_0 (r-1)\right] W_0(q)\right\}&
	-q^2 D_0\left\{1-2\chi N_p\left[\tilde{m}_2-\tilde{m}_1^2 (1-(r-1) (1-\tilde{m}_0))\right]W_0(q)^2\right\} \end{array}\right)
\end{equation}
\normalsize

 Using Mathematica, we get its corresponding eigenvalues

\begin{eqnarray}
\lambda_0&=&\frac{D_0 q^2}{2}\left\{-1-(1+\tilde{m}_0(r-1))^2+2\chi\left[\tilde{m}_2+\tilde{m}_1^2(-1+(1-\tilde{m}_0)(r-1))\right]+ \Delta\right\}\nonumber\\
&+& \frac{D_0 q^4}{48}\left\{4 \chi (-\tilde{m}_2 + \tilde{m}_1^2 (2 + \tilde{m}_0 (-1 + r) - r)) + 
    \tilde{m}_0 (1 + \tilde{m}_0 (-1 + r)) r \right\}\nonumber\\
&+& \frac{D_0 q^4}{48\Delta}\left\{\tilde{m}_0^2 (1 + \tilde{m}_0 (-1 + r)) (2 + \tilde{m}_0 (-1 + r)) (-1 + r) r + 
      8 \chi^2 (\tilde{m}_2 + \tilde{m}_1^2 (-2 + \tilde{m}_0 + r - \tilde{m}_0 r))^2\right.\nonumber\\
&+&\left. 
      2 \chi (-4 + 5 r + \tilde{m}_0 (-1 + r) (-2 + 3 r)) 
       (\tilde{m}_0 \tilde{m}_2 + \tilde{m}_1^2 (-2 + \tilde{m}_0 (2 - \tilde{m}_0 + (-1 + \tilde{m}_0) r)))\right\}\\
\lambda_1&=& \frac{D_0 q^2}{2}\left\{-1-(1+\tilde{m}_0(r-1))^2+2\chi\left[\tilde{m}_1^2(r-1)(1-\tilde{m}_0)+\tilde{m}_2-\tilde{m}_1^2\right]-  \Delta\right\}\nonumber\\
  &+&\frac{D_0 q^4}{48}\left\{4 \chi (-\tilde{m}_2 + \tilde{m}_1^2 (2 + \tilde{m}_0 (-1 + r) - r)) + 
    \tilde{m}_0 (1 + \tilde{m}_0 (-1 + r)) r \right\}\nonumber\\
&+&\frac{D_0 q^4}{48\Delta}\left\{
    \-(\tilde{m}_0^2 (1 + \tilde{m}_0 (-1 + r)) (2 + \tilde{m}_0 (-1 + r)) (-1 + r) r) - 
      8 \chi^2 (\tilde{m}_2 + \tilde{m}_1^2 (-2 + \tilde{m}_0 + r - \tilde{m}_0 r))^2 \right.\nonumber\\
&+&\left.- 
      2 \chi (-4 + 5 r + \tilde{m}_0 (-1 + r) (-2 + 3 r)) 
       (\tilde{m}_0 \tilde{m}_2 + \tilde{m}_1^2 (-2 + \tilde{m}_0 (2 - \tilde{m}_0 + (-1 + \tilde{m}_0) r)))\right\}
\end{eqnarray}
where the quantity  $\Delta$ can be expressed as
\begin{eqnarray}
\Delta&\equiv&
\left\{(1+(1+\tilde{m}_0(r-1))^2+2\chi\left[-\tilde{m}_2+\tilde{m}_1^2-\tilde{m}_1^2(r-1)(1-\tilde{m}_0)\right])^2\right.\nonumber\\
&+&\left. 4
(1+\tilde{m}_0(r-1))(-1-\tilde{m}_0(r-1)+2\chi\left[\tilde{m}_2(1+\tilde{m}_0(r-1))-\tilde{m}_1^2r\right])\right\}^{1/2}
\end{eqnarray}

\section{General expression for fluxes in length-polydisperse case}
\label{app:2}
In the main text we have given only the
expressions for the diffusive fluxes in the linearized case because the
generic expressions become quite lengthy, and display them here
instead.  The
generic expressions for the diffusive flux of species $i$ can be
expressed as
\begin{eqnarray}
\alpha {\bf J}_i &=& -\frac{1}{N_i^2}\nabla\rho_i+\rho_i \nabla
m_{-2}+\rho_i\left[\frac{1}{N_i}-\left(m_{-1}+\frac{1-m_0}{N_s}\right)-
\frac{1}{24}\nabla^2+\frac{1}{24} \nabla^2 m_0\right]\nabla m_{-1}\nonumber\\
&+&\left\{2 \rho_i\chi\left[\frac{W_{0i}}{N_i}-\frac{1}{N_i}(m_0-\frac{1}{24}\nabla^2m_1+\frac{1}{24}m_1\nabla^2)-m_{-1}+\frac{1}{24}\nabla^2m_0-\frac{1}{24}m_0\nabla^2\right.\right.
\nonumber\\
&+&\left.\left(m_{-1}-\frac{1}{24}\nabla^2m_0+\frac{1-m_0}{N_s}\right)\left(m_0-\frac{1}{24}\nabla^2m_1+\frac{1}{24}m_1\nabla^2\right)\right]\nonumber\\
&+&\rho_i\left[-\frac{1}{24 N_i}\nabla^2+\left(m_{-1}+\frac{1-m_0}{N_s}\right)\frac{1}{24}\nabla^2\right.
\nonumber\\
&-&\left.\left.\frac{1}{N_s}\left(\frac{1}{N_i}+\frac{m_0}{N_s}-m_{-1}+\frac{1}{24}\nabla^2m_0\right)\right]\right\}\nabla m_0
\end{eqnarray}
while the corresponding fluxes for the moments can be expressed as      
\begin{eqnarray}
{\bf j}_n &=& -\nabla m_{n-2}+m_n\nabla
m_{-2}+\left[m_{n-1}-m_nm_{-1}-\frac{m_n (1-m_0)}{N_s}+\frac{1}{24} m_n(\nabla^2
m_0-\nabla^2)\right]\nabla m_{-1} \nonumber\\
&+&\left\{2\chi\left[(1-m_0)\left(m_{n-1}-m_nm_{-1}+\frac{m_nm_0}{N_s}\right)+\frac{1}{24}\left[\nabla^2
m_n+m_n ((1-m_0)
\nabla^2m_0-m_0\nabla^2)
\right.\right.\right.\nonumber\\
&+&\left.\left.\left(m_nm_{-1}+\frac{m_n(1-m_0)}{N_s}-m_{n-1}\right)(m_1\nabla^2-\nabla^2m_1)\right]\right]-\frac{m_{n-1}-m_n(m_{-1}-m_0/N_s)}{N_s}\nonumber\\
&+&\frac{1}{24}\left.\left(-m_{n-1}+m_nm_{-1}+\frac{m_n
(1- m_0)}{N_s}\right)\nabla^2-\frac{m_n}{24 N_s}\nabla^2m_0\right\}\nabla m_0
\end{eqnarray}

\section{Diffusive fluxes after dynamical projection}
\label{app:3}
We provide here the expression for the diffusive fluxes of the
moments $m_{-2}$, $m_{-1}$ and $m_0$, when we have derived  a closed set
of equation for their dynamics. These expressions were used to obtain numerically the three corresponding eigenmodes, and can be written down as
\begin{eqnarray}
{\bf j}_{-2} &=& \left(-\langle
N\rangle^2\gamma_{\bar{4}\bar{2}}+\tilde{m}_0\langle N\rangle\langle
N^{-1}\rangle\right)\frac{1}{\langle
N\rangle^2}\nabla m_{-2} \nonumber\\
&+&\left[-\gamma_{\bar{4}\bar{1}}\langle
N\rangle^3+\tilde{m}_0\langle N^{-2}\rangle\langle N\rangle^2-\tilde{m}_0\langle
N^{-1}\rangle\langle N\rangle(r(1-\tilde{m}_0)+\tilde{m}_0)\right]\frac{\nabla m_{-1}}{\langle N\rangle^3}\nonumber\\
&+&\left[-\langle
N\rangle^4\gamma_{\bar{4}0}+\tilde{m}_0(2\chi \langle N\rangle
(1-\tilde{m}_0)-r)(\langle N^{-2}\rangle\langle N\rangle^2+\tilde{m}_0(r-1)\langle
N^{-1}\rangle\langle N\rangle)\right]\frac{\nabla
m_0}{\langle N\rangle^4}\nonumber\\
{\bf j}_{-1} &=& (-\langle
N\rangle\gamma_{\bar{3}\bar{2}}+\tilde{m}_0) \frac{1}{\langle
N\rangle}\nabla m_{-2}+(-\langle
N\rangle^2\gamma_{\bar{3}\bar{1}}+\tilde{m}_0\langle N^{-1}\rangle\langle
N\rangle-\tilde{m}_0((1-\tilde{m}_0) r+\tilde{m}_0))\frac{1}{\langle N\rangle^2}\nabla
m_{-1}\nonumber\\
&+&\left[-\langle N\rangle^3\gamma_{\bar{3}0}+2\chi \langle N\rangle
\tilde{m}_0 (1-\tilde{m}_0)(\langle N^{-1}\rangle\langle N\rangle+\tilde{m}_0 (r-1))-r \tilde{m}_0
(\langle N^{-1}\rangle\langle N\rangle+\tilde{m}_0 (r-1))\right]\frac{1}{\langle N\rangle^3}\nabla m_0\nonumber\\
{\bf j}_0 &=&-(1-\tilde{m}_0)\nabla m_{-2}-\tilde{m}_0 (1-\tilde{m}_0)(r-1)\frac{\nabla m_{-1}}{\langle
N\rangle}\nonumber\\
&+&m_0\left[2\chi\langle N\rangle(1-\tilde{m}_0)(1+\tilde{m}_0 (r-1))-r(1+\tilde{m}_0(r-1)) \right]\frac{\nabla m_0}{\langle N\rangle^2}
\end{eqnarray}

The coefficients $\gamma_{\bar{i}\bar{j}}$ with $i=4,3$ and $j=2,1,0$
are the equivalent of the coefficients $\beta_{\bar{i}\bar{j}}$ in
eqs \ref{eq:projection} when we project the moments $m_{-4}$ and $m_{-3}$
into the subspace spanned by the moments $m_{-2}$, $m_{-1}$ and
$m_0$. The values of these coefficients are functions of the subspace
chosen.

\newpage

{\bf Figure captions}

\begin{enumerate}

\item { {\bf Figure 1:} Wavevector dependence of the unstable eigenmode $\lambda_1 N_p/D_0$
 for an exponential parent $\rho_n\sim\exp (a f_n)$ at $\chi=1.2
\chi_s$. The wave vector $q$  is
measured in units of $N_p^{-1/2}$. In all the plots, we show the
eigenmodes for four different size ratios $r=10,50,100,300$ and $1200$. In the
different plots we change the degree of asymmetry of the chemical
distribution (increasing with increasing $a$), and compare dense and
dilute suspensions. a) $a=2$, $\tilde{m}_0=0.15$; b) $a=5$, $\tilde{m}_0=0.15$; c) $a=10$, $\tilde{m}_0=0.15$ ; d) $a=10$, $\tilde{m}_0=0.8$.}

\item { {\bf Figure 2:} Wavevector dependence of the stable eigenmode $\lambda_0 N_p/D_0$
 for an exponential parent $\rho_n\sim\exp (a f_n)$ at high dilution
$m_0=0.15$ and $\chi=1.2 \chi_s$. The wave vector $q$ is
measured in units of $N_p^{-1/2}$.  a) $a=2$; b) $a=10$.}

\item {{\bf Figure 3:} Quenched (long-dashed) and annealed (continuous) spinodals for a chemically \\ polydisperse
mixture with an exponential parent shape $\rho_n\sim \exp (a f_n)$ as a
function of monomer concentration for
various degrees of asymmetry $a$ and size ratio $r$. The interaction
strength $\chi$ is normalized by its value at the critical point, $\chi_c$,
for the corresponding set of parameters $r$ and $a$. a) $a=1$, $r=10$;
b)  $a=5$, $r=10$; c)  $a=20$, $r=10$;d)  $a=20$, $r=1000$.}
\end{enumerate}
\begin{center}
\begin{figure}[tbh]
\epsfig{figure=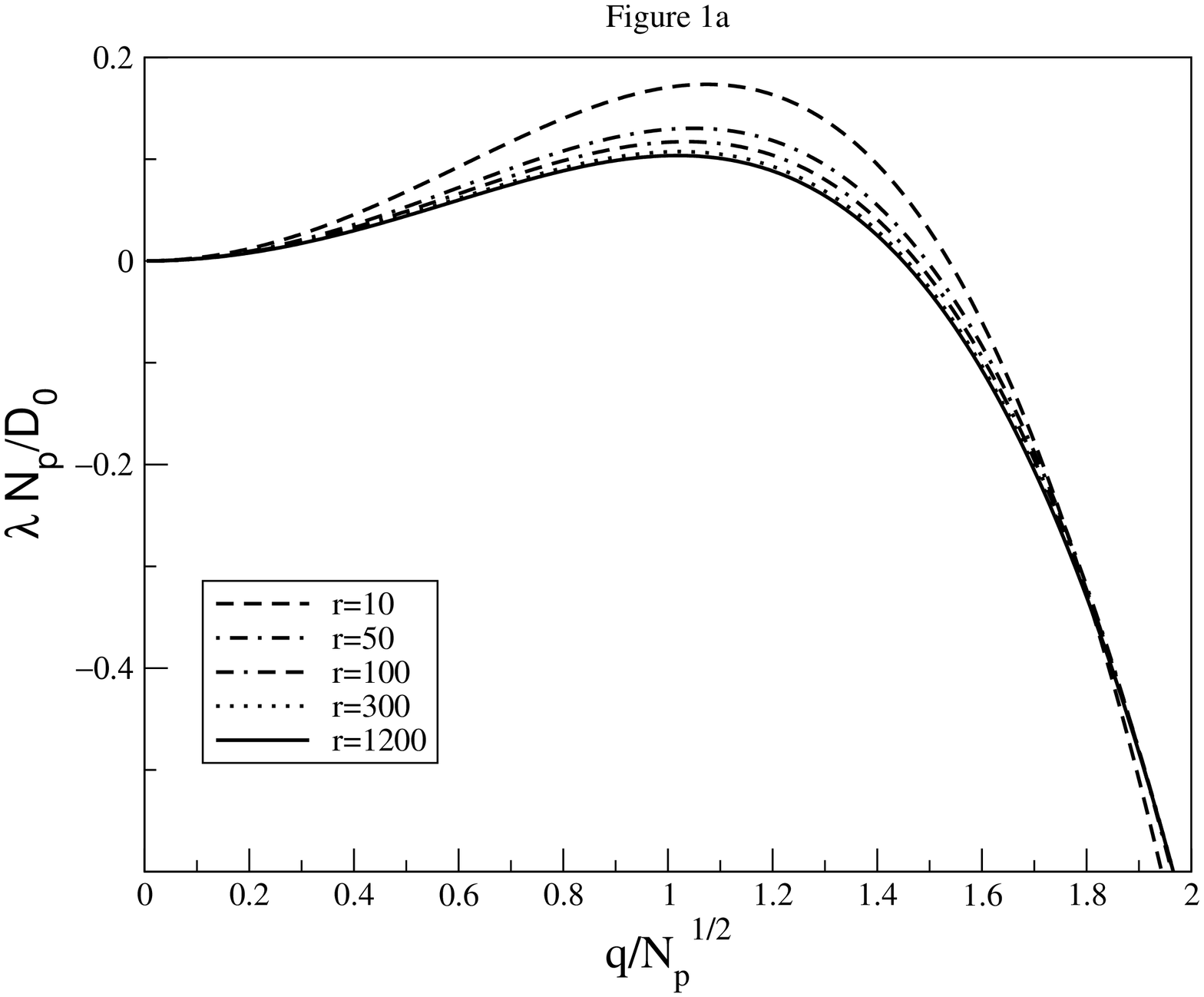,width=16cm,angle=-90}
\end{figure}
\begin{figure}[tbh]
\epsfig{figure=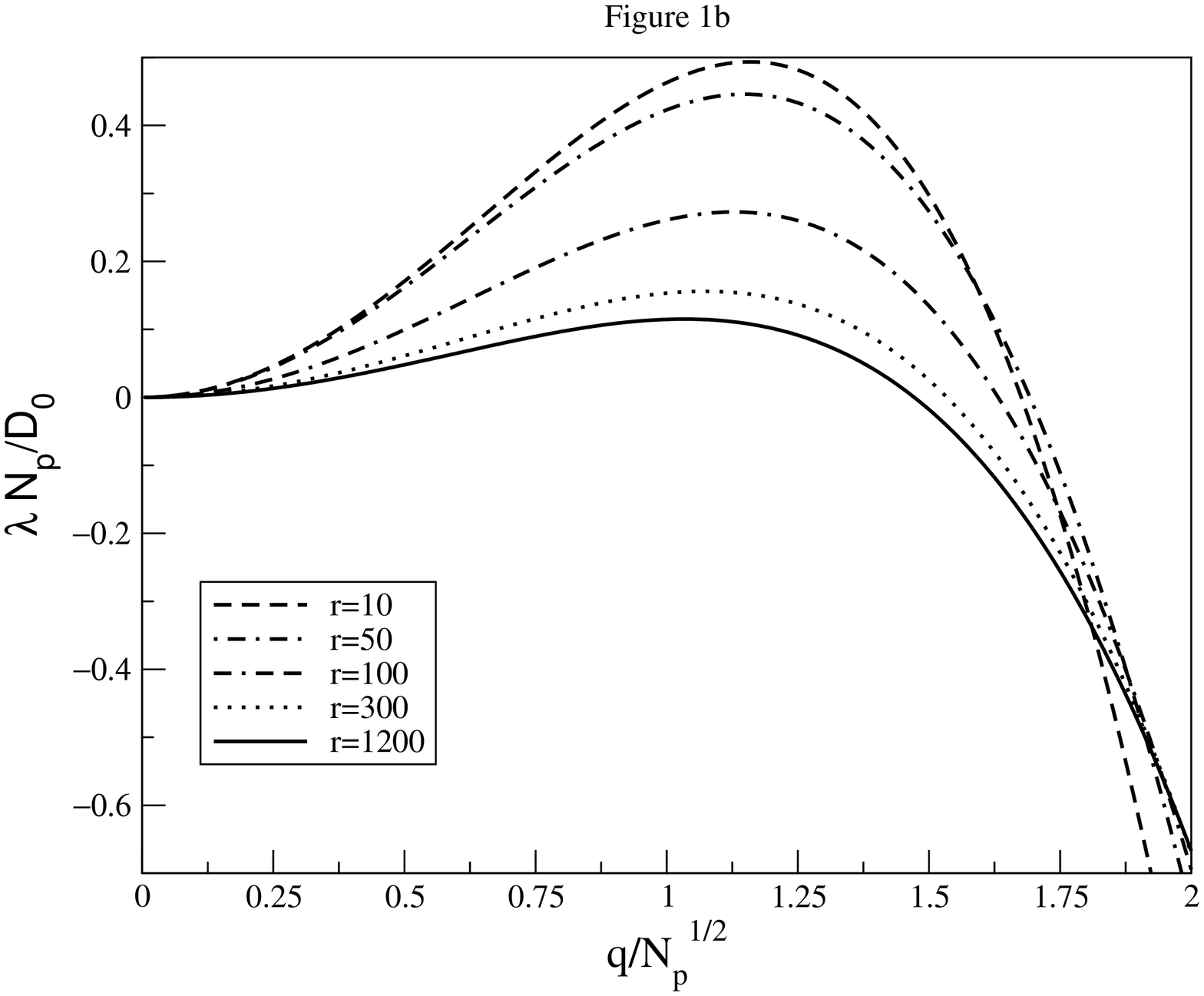,width=16cm,angle=-90}
\end{figure}
\begin{figure}[tbh]
\epsfig{figure=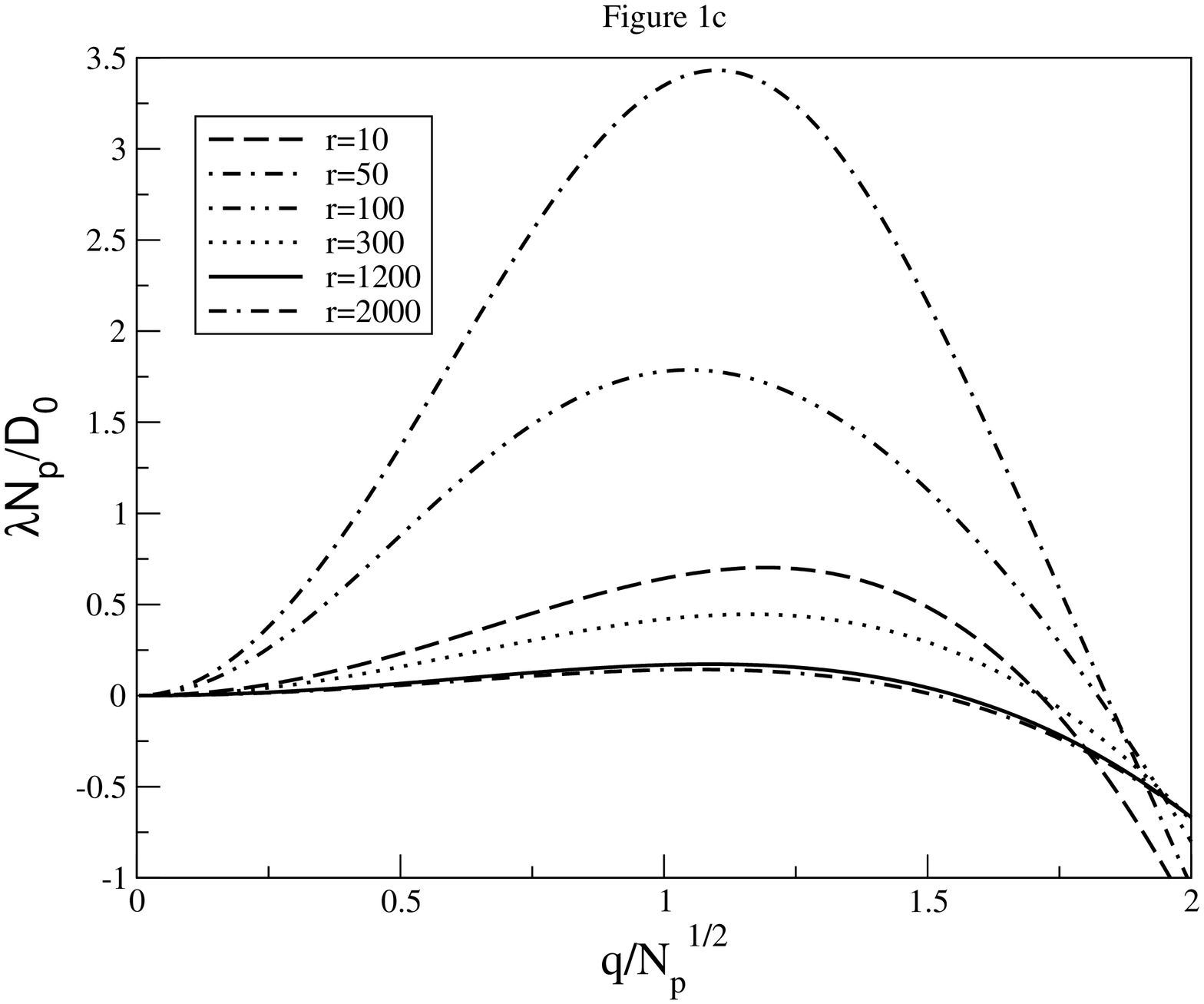,width=16cm,angle=-90}
\end{figure}
\begin{figure}[tbh]
\end{figure}
\begin{figure}[tbh]
\epsfig{figure=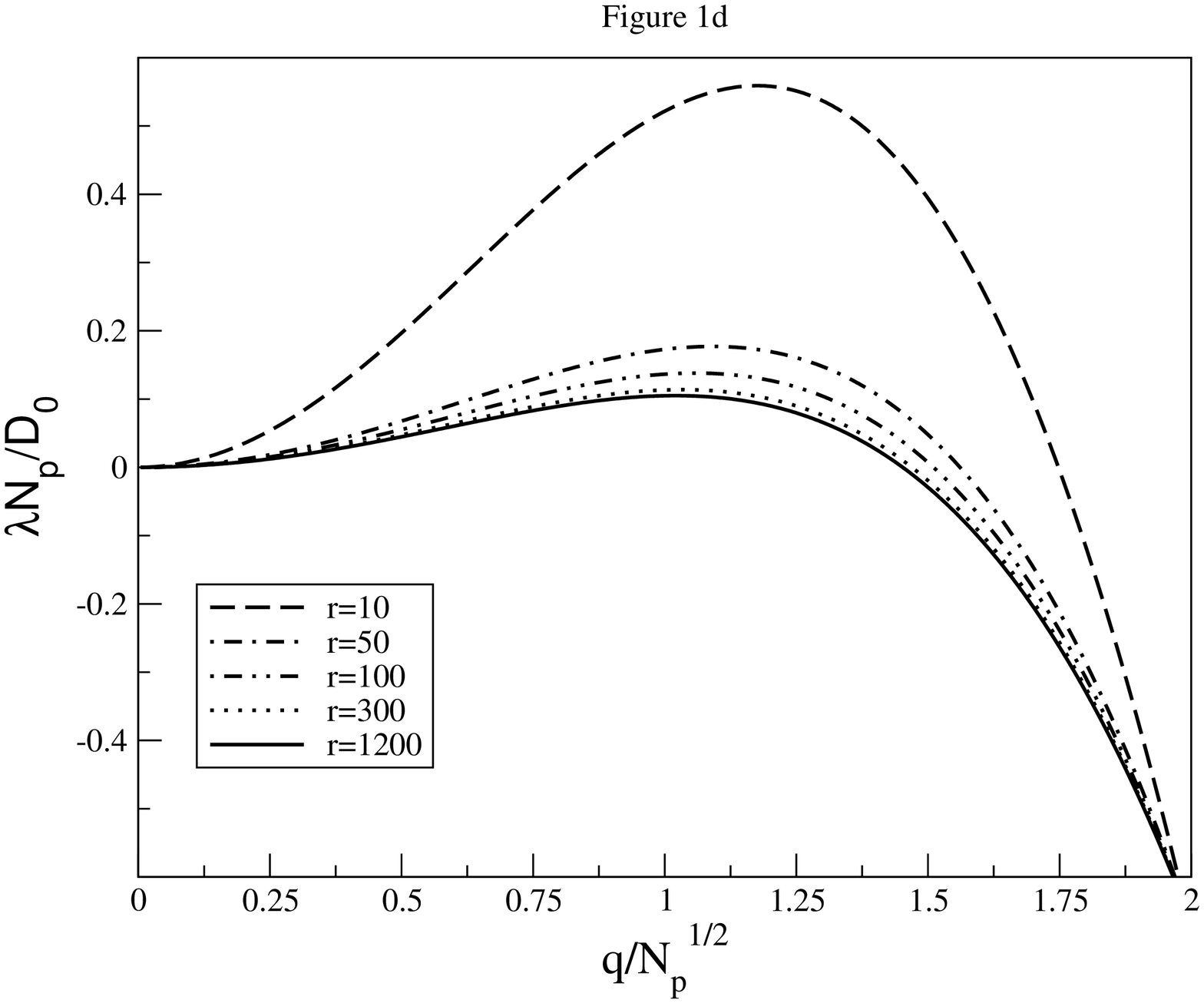,width=16cm,angle=-90}
\end{figure}
\begin{figure}[tbh]
\epsfig{figure=./fig2a.ps,width=16cm,angle=-90}
\end{figure}
\begin{figure}[tbh]
\epsfig{figure=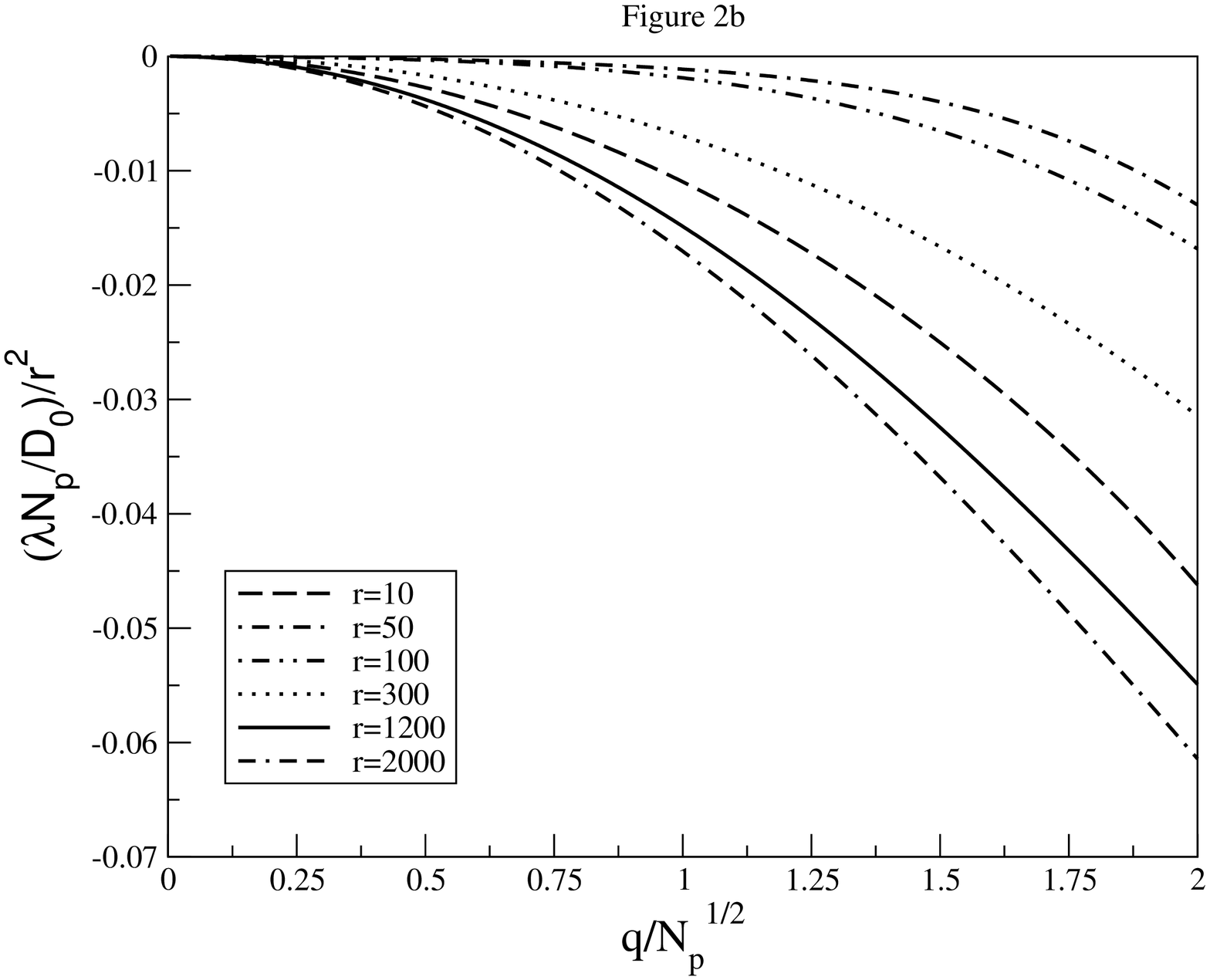,width=16cm,angle=-90}
\end{figure}
\begin{figure}[tbh]
\epsfig{figure=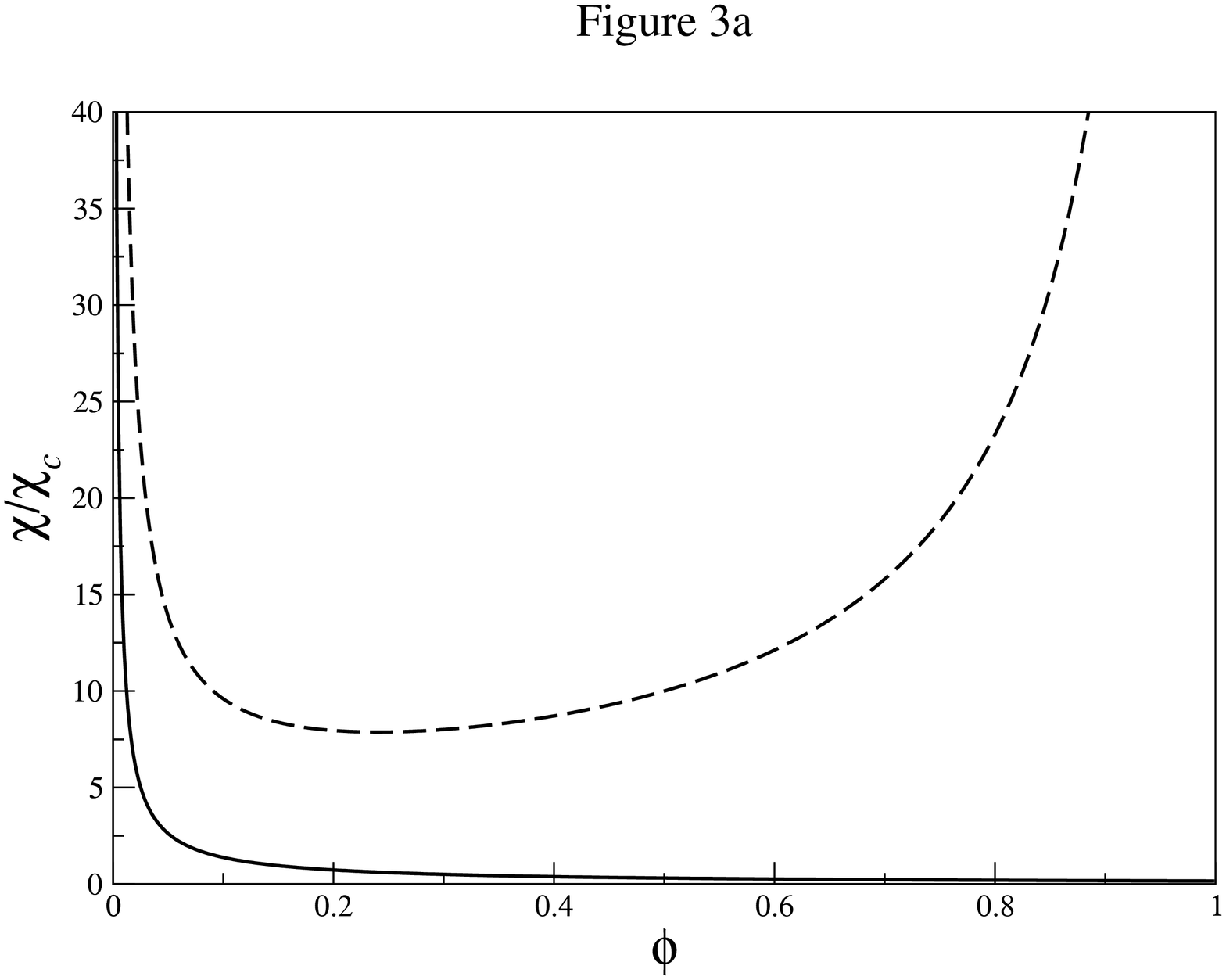,width=16cm,angle=-90}
\end{figure}
\begin{figure}[tbh]
\epsfig{figure=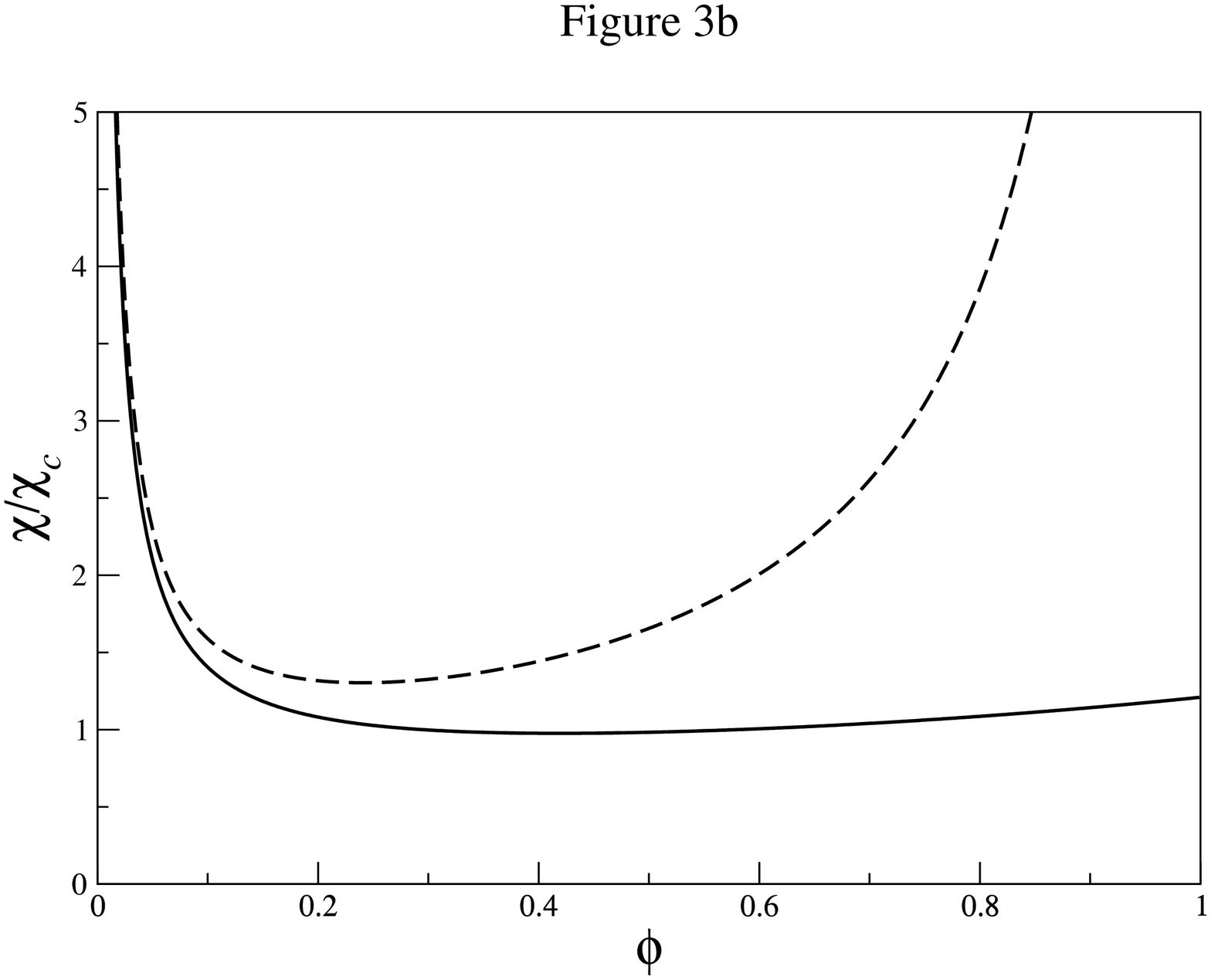,width=16cm,angle=-90}
\end{figure}
\begin{figure}[tbh]
\epsfig{figure=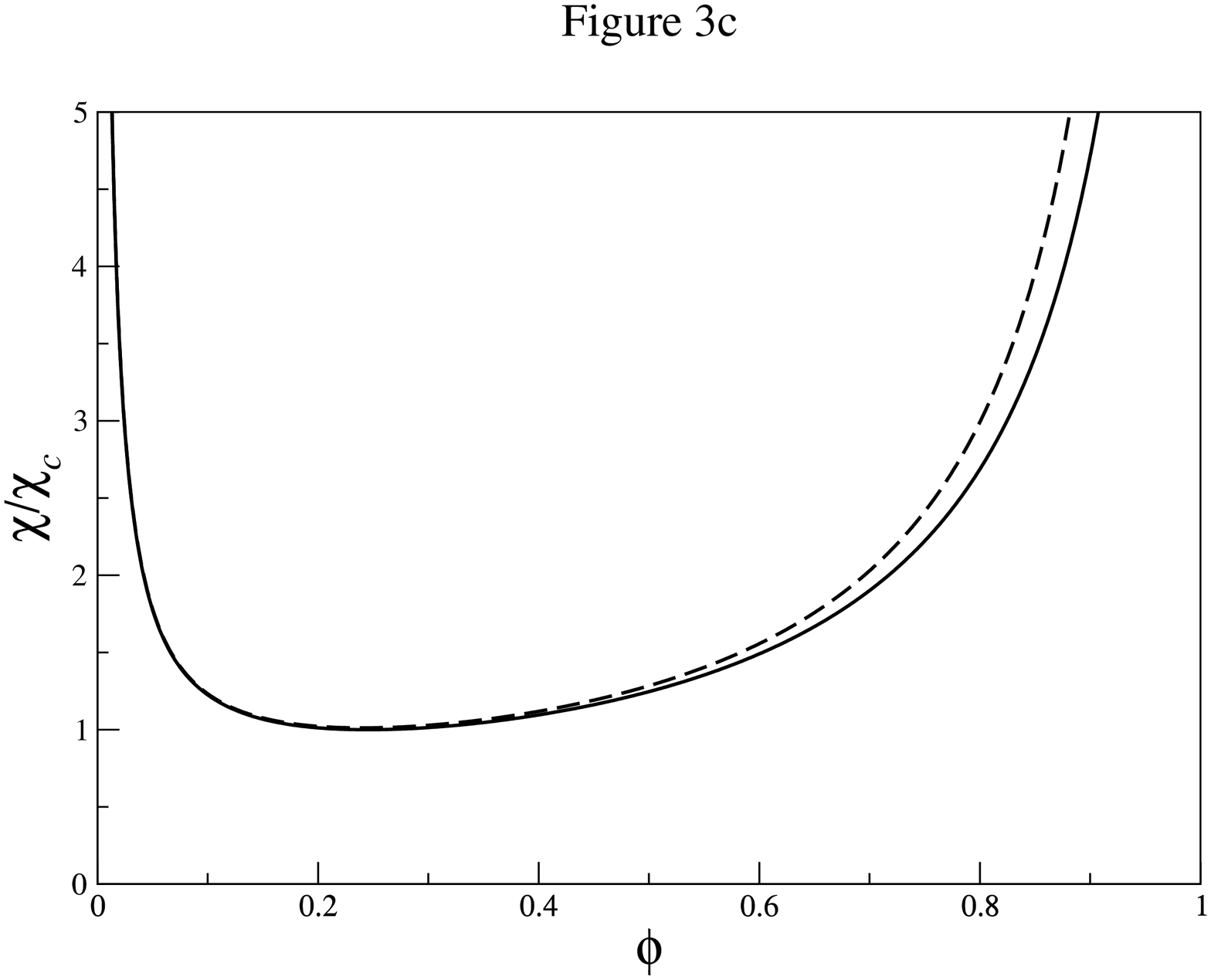,width=16cm,angle=-90}
\end{figure}
\begin{figure}[tbh]
\epsfig{figure=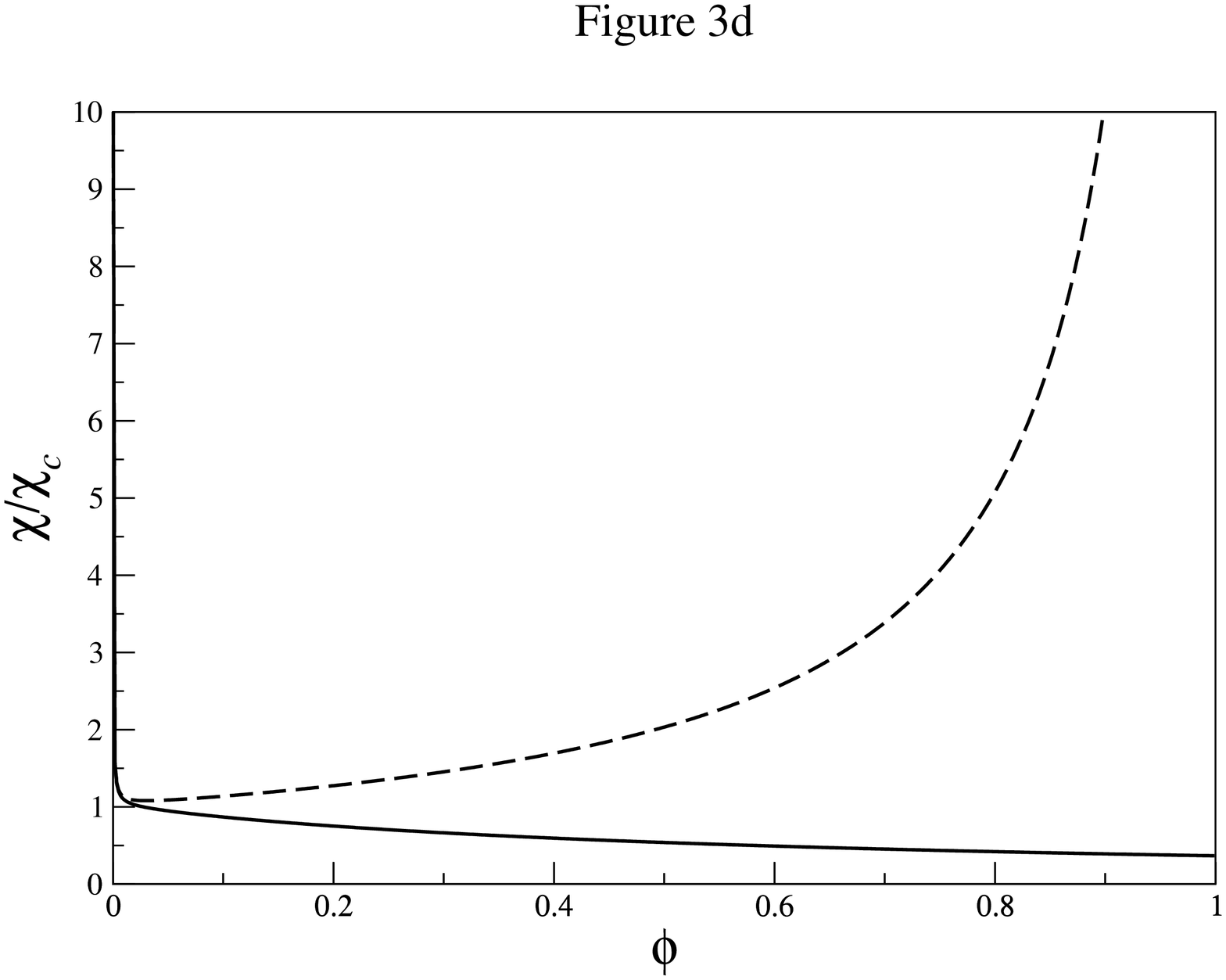,width=16cm,angle=-90}
\end{figure}
\end{center}

\begin{thebibliography}{99}
\bibitem{footgD} The weight function is related to the square root of
the Debye function. For a polymer of length $N_i$, one has  $\omega(q)=N_i
g(q)^{1/2}$. In the  text we are interested in the low $q$ 
expansion of the Debye function $g(q)$; see ref 2. $g(q)$ also
depends on the polymer length through the gyration radius.
\bibitem{footrates} Note that the eigenvalues $\lambda_i$ are rates that control 
the initial growth of the moments. They are then characteristic frequencies 
defined by $\delta m_n(t)=\delta m_n(0)\exp(\lambda_n t)$.
\bibitem{footquasitri} Except for the fact that $m_0$ depends on $m_1$, the 
matrix of coefficients that defines the linear dynamics of this system is lower triangular.
\bibitem{notation} What we call
$\tilde{\lambda}_i$ (with $i$ being a species) corresponds to the
coefficients  $\lambda_i$ in ref 4. We have introduced the tilde 
to avoid confusion with the growth rates, which are referred to as  $\lambda_n$ 
(with $n$ being a moment).
\bibitem{footfast} The fact that
Brochard's model is a fast mode theory can be appreciated in the
expressions for the diffusive fluxes, eq \ref{eq:mik}. The
contribution of species $j$ to the flux of species $i$  is
inverserly proportional to its friction coefficient (which in turn is
proportional to its  polymerization length). Hence, the
longer the chain the smaller the cooresponding contribution to the
diffusive flux of the rest of the species.
\end{thebibliography}
\end{document}